\documentclass[prd,reprint,twocolumn]{revtex4-1}

\usepackage{amssymb}
\usepackage{amsfonts}
\usepackage{amstext}
\usepackage{graphicx}
\usepackage{amscd}
\usepackage{amsmath}
\usepackage{mathrsfs}
\usepackage{stmaryrd}
\usepackage{slashed}

\newcommand{\Sp}{\mathrm{Sp}}

\renewcommand{\Re}{\mathrm{Re}}
\renewcommand{\Im}{\mathrm{Im}}

\newcommand{\llangle}{\langle \hspace{-0.2em} \langle}
\newcommand{\rrangle}{\rangle \hspace{-0.2em} \rangle}
\newcommand{\lllangle}{\langle \hspace{-0.2em} \langle \hspace{-0.2em} \langle}
\newcommand{\rrrangle}{\rangle \hspace{-0.2em} \rangle \hspace{-0.2em} \rangle}

\newcommand{\id}{\mathrm{id}}

\newcommand{\dnc}{\mathbf{d}}

\newcommand{\Lin}{\mathrm{Lin}}
\newcommand{\tr}{\mathrm{tr}}

\begin{document}

\title{Fuzzy Schwarzschild (2+1)-spacetime}

\author{David Viennot}
\affiliation{Institut UTINAM (CNRS UMR 6213, Universit\'e de Franche-Comt\'e), 41bis Avenue de l'Observatoire, BP 1615, 25010 Besan\c con cedex, France.}
\email{david.viennot@utinam.cnrs.fr}

\begin{abstract}
  We present a toy model of fuzzy Schwarzschild space slice (as a noncommutative manifold) which quantum mean values and quantum quasi-coherent states (states minimizing the quantum uncertainties) have properties close to the classical slice of the $(r,\theta)$ Schwarzschild coordinates (the so-called Flamm's paraboloid). This fuzzy Schwarzschild slice is built as a deformation of the noncommutative plane. Quantum time observables are introduced to add a time quantization in the model. We study the structure of the quasi-coherent state of the fuzzy Schwarzschild slice with respect to the quasi-coherent state and the deformation states of the noncommutative plane . The quantum dynamics of a fermion interacting with a fuzzy black hole described by the present model is studied. In particular we study the decoherence effects appearing in the neighborhood of the fuzzy event horizon.  An extension of the model to describe a quantum wormhole is also proposed, where we show that fermions cross the wormhole not by traveling by its internal space but by quantum tunneling, in accordance with the non-traversable character of classical Einstein-Rosen bridges.
\end{abstract}

\maketitle

\section{Introduction}
Fuzzy manifolds \cite{Berenstein, Steinacker, Schneiderbauer, Sykora} are simple examples of noncommutative manifolds, which can be used as quantum gravity models by supposing a spacetime quantization based on the Connes' noncommutative geometry \cite{Connes}. These models can be also interpreted as matrix models in string theory. Basically, in $\mathbb R^3$, we can define a fuzzy manifold $\mathscr M$ by quantization of the coordinate observables $(x,y,z)$ of a classical manifold, which become quantum observable operators $(X,Y,Z)$, generating the algebra of the ``noncommutative functions'' of $\mathscr M$ and satisfying a relation similar to the embedding equation for $(x,y,z)$. Two simple examples are the noncommutative plane and the fuzzy sphere. A classical plane being defined by the embedding equation $z=0$, a non-commutative plane can be defined by
\begin{equation}\label{NCPlane}
  X = \frac{a+a^+}{2}, \qquad Y=\frac{a-a^+}{2\imath}, \qquad Z=0
\end{equation}
where $a$ and $a^+$ are the quantum harmonic oscillator annihilation and creation operators ensuring that $[X,Y]=\frac{\imath}{2} \not=0$. A classical sphere being defined by the embedding equation $x^2+y^2+z^2=r^2$, a fuzzy sphere can be defined by
\begin{equation}
  X^2+Y^2+Z^2 = \lambda^2j(j+1)
\end{equation}
(with $\lambda >0$ a parameter) implying that $(X/\lambda,Y/\lambda,Z/\lambda)$ are the generators of the $(2j+1)$ dimensional irreducible representation of the quantum angular momentum algebra. In this paper, we are interested by fuzzy manifolds mimicking the Schwarzschild geometry of a neutral non-rotating black hole, as a model of quantum micro black hole \cite{Hawking,Hawking2, Giddings, Calmet}.\\
Various approaches have been used to built quantum black hole models from noncommutative geometry: by identifying the event horizon with a fuzzy sphere \cite{Dolan,Iizuka,Argota}; by using a Moyal-type star product with an embedding geometry in $\mathbb R^7$ \cite{Blaschke}, or by considering a classical geometry modified by a non-commutative parameter \cite{Nicolini, Banerjee}. In this paper we propose another approach with a toy model based on a deformation of the noncommutative plane in order to its mean values in the quasi-coherent state (the quantum state closest to a classical one) corresponds to the classical embedding in $\mathbb R^3$ of a ``space slice'' of the Schwarzschild geometry. The quantization of time observables is then obtained from an analysis of the dynamics of the model. Such a model permits to study the quantum dynamics of a fermionic particle in the neighborhood of the quantum black hole in the quasi-coherent representation.\\

This paper is organized as follows. Section II presents the general theoretical framework used in this paper: fuzzy manifolds, their Dirac operators and the notion of quasi-coherent states. Section III introduces the model of fuzzy Schwarzschild slice and introduces associated quantum time observables. The Dirac operator is not self-adjoint and includes a dissipator which can be interpreted as modeling the particle absorption by the black hole singularity.  Section IV studies the quasi-coherent states of the fuzzy Schwarzschild slice (especially the classical geometry closest to the fuzzy geometry). Quantum dynamics governed by the Dirac operator of the fuzzy Schwarzschild spacetime are studied section V. The evolution of the quantum coherence of the states is studied and the mean value and the quasi-coherent wave packet of the quantum dynamics are compared with the classical dynamics of a probe D0-brane (viewed as a classical point particle and driving the quantum evolution). Section VI extends the present model to white holes and to structures like wormholes. We show in this section, that a particle crossing the wormhole do not travel by its internal space but crosses it by quantum tunneling. Appendix A presents the case of a simple noncommutative plane to compare with the fuzzy Schwarzschild slice and Appendix B presents the numerical method used for the computations.\\

\textit{Throughout this paper, we consider the unit system such that $\hbar=c=G=1$ ($\ell_P=t_P=m_P=1$ for the Planck units).}

\section{Noncommutative Dirac operator and quasi-coherent states}
First we recall some basic facts about fuzzy manifolds.\\

Let $(X,Y,Z)$ be three coordinate self-adjoint operators defining the fuzzy manifold $\mathscr M$. We denote by $\mathfrak a$ the algebra generated by $(X,Y,Z)$ (universal enveloping algebra of the algebra of the polynomials of $X$, $Y$ and $Z$). $\mathfrak a$ plays the role of a set of non-abelian functions of $\mathscr M$. Let $\mathscr F$ be a Hilbert space of representation of $\mathfrak a$. The fundamental operator of the noncommutative geometry is the following Dirac operator \cite{Berenstein, Steinacker, Schneiderbauer, Sykora}
\begin{equation}
  \slashed D_{\alpha,z} = \left(\begin{array}{cc} Z-z & A^+ - \bar \alpha \\ A-\alpha & -Z+z \end{array}\right) \in \mathcal L(\mathbb C^2 \otimes \mathscr F)
\end{equation}
where $A=X+\imath Y$ and $A^+=X-\imath Y$, and $(\alpha,z)\in \mathbb C \times \mathbb R$. The Hilbert space $\mathscr H = \mathbb C^2 \otimes \mathscr F$ is the space of spinors on $\mathscr M$ ($\mathbb  C^2$ is the spin state space). $(\mathfrak a,\mathscr H,\slashed D_{\alpha,z})$ is a spectral triple of the Connes' theory \cite{Connes}. $(\Re\, \alpha,\Im\, \alpha,z) \in \mathbb R^3$ is a point of the embedding space. $(\alpha,z)$ are classical control parameters of $\slashed D_{\alpha,z}$. In the context of string theory, in the BFFS matrix model \cite{Banks}, $(X,Y,Z)$ can be viewed as the operators of a stack of D0-branes and $\slashed D_{\alpha,z}$ is the Dirac operator of a fermionic string linking this stack to a probe D0-brane of coordinates $(\Re\, \alpha,\Im\, \alpha,z)$ in the embedding space \cite{Berenstein, Steinacker} (the spacetime being reduced to 3+1 dimensions by a truncation with a supersymmetric orbifold, see \cite{Berenstein}). $\mathscr F$ plays the role of the Hilbert space of ``position'' states on $\mathscr M$ of the end of the fermionic string. The string spinor $|\psi \rrangle \in \mathscr H$ obeys to the noncommutative Dirac equation \cite{Berenstein, Steinacker}
\begin{equation}
  \imath |\dot \psi \rrangle = \slashed D_{\alpha(\tau),z(\tau)} |\psi(\tau) \rrangle
\end{equation}
where the dot denotes the derivative with respect to $\tau$. This equation corresponds to the dynamics of the fermionic string induced by the transport in the embedding space of the probe D0-brane described by $\tau \mapsto (\alpha(\tau),z(\tau))$. Consistently, $\tau$ is defined as the proper time of an observer comoving with the probe D0-brane.\\

To compare the geometry of $\mathscr M$ to a classical geometry, we need to find a quantum state of $\mathscr M$ closest to a classical state. For quantum systems described by a Lie algebra, the quantum states closest to classical ones are the Perelomov coherent states \cite{Perelomov}, for which the Heisenberg uncertainty relations are minimized. The equivalent notion for a fuzzy manifold $\mathscr M$, called quasi-coherent states, are the eigenvectors of $\slashed D_{\alpha,z}$ in its kernel \cite{Schneiderbauer}:
\begin{equation}\label{Diraceq1}
  \slashed D_{\alpha,z}|\Lambda(\alpha,z) \rrangle = 0
\end{equation}
If $(X,Y,Z)$ generates a Lie algebra, $|\Lambda(\alpha,z)\rrangle$ is strongly related to the Perelomov coherent state (see the example of the noncommutative plane in appendix \ref{NCP}). More generally, ref. \cite{Schneiderbauer} shows that $|\Lambda(\alpha,z)\rrangle$ minimizes the \textit{displacement energy} which can be viewed as the ``tension energy'' of the fermionic string. This one is large if the probe D0-brane is moved away $\mathscr M$ and/or if the dispersion $\Delta X^2 + \Delta Y^2 + \Delta Z^2$ is large (with $\Delta X^2 = \llangle X^2 \rrangle - \llangle X \rrangle^2$). This means that, from the point of view of the fermionic string, $|\Lambda \rrangle$ is the state for which the end of the string is as less delocalized as possible and for which the probe D0-bane (the other end) is closest to this end. In other words, $|\Lambda \rrangle$ is the quantum state of the fermionic string closest to a state of a classical point particle. The values of $(\alpha,z)$ for which eq.(\ref{Diraceq1}) has solutions form a classical manifold in the embedding space:
\begin{equation}
  M_\Lambda = \{(\alpha,z)\in \mathbb C\times \mathbb R, \text{ s.t. } \det \slashed D_{\alpha,z} = 0 \} \subset \mathbb R^3
\end{equation}
We call $M_\Lambda$ the eigenmanifold of $\mathscr M$, and since \cite{Schneiderbauer}
\begin{eqnarray}
  & & \forall (\alpha,z) \in M_\Lambda, \nonumber \\
  & & \quad \llangle \Lambda(\alpha,z)|A|\Lambda(\alpha,z)\rrangle = \alpha \\
  & & \quad \llangle \Lambda(\alpha,z)|A^+|\Lambda(\alpha,z)\rrangle = \bar \alpha \\
  & & \quad \llangle \Lambda(\alpha,z)|Z|\Lambda(\alpha,z)\rrangle = z
\end{eqnarray}
We can see $M_\Lambda$ as the ``mean value'' of $\mathscr M$. $M_\Lambda$ is then the classical manifold closest to the fuzzy manifold $\mathscr M$, in the sense that the mean values of the coordinates operators of $\mathscr M$ in the quasi-coherent states are coordinates of points of $M_\Lambda$, with minimal quantum uncertainty $\Delta X^2+\Delta Y^2+\Delta Z^2$. For a point of view of quantum gravity, if $\mathscr M$ describes a quantum space, $M_\Lambda$ is the corresponding slice of a curved classical space revealed by the transport of the test point particle equivalent to the fermionic string in the quasi-coherent state.\\

The goal of this paper is to present a toy model of a fuzzy manifold for which its eigenmanifold is close to the Schwarzschild geometry.

\section{The model of fuzzy Schwarzschild spacetime}
\subsection{Coordinate operators}
The embedding of a space slice of the Schwarzschild geometry is defined (in Schwarzschild coordinates) by
\begin{equation}\label{SchwarzEq}
  z = f(r) = 2 \sqrt{r_S} \sqrt{r-r_S}
\end{equation}
where $r_S$ is the Schwarzschild radius. This embedding induces then the spatial part of the Schwarzschild metric
\begin{equation}
  (1+f'(r)^2) dr^2+r^2d\theta^2 = \left(1-\frac{r_S}{r}\right)^{-1} dr^2 + r^2 d\theta^2
\end{equation}
(by setting that $\mathbb R^3$ is endowed with its canonical cartesian metric).\\

\begin{figure}
  \includegraphics[width=8cm]{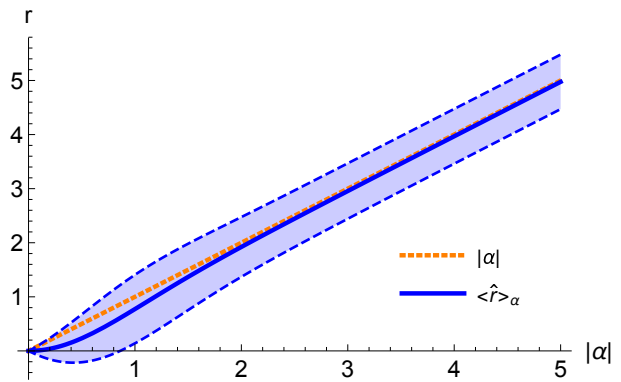}\\
  \caption{\label{rmoy} $\langle \hat r \rangle_\alpha = \langle \alpha|\hat r|\alpha \rangle$ with respect to $|\alpha|$. The dashed lines correspond to $\langle \hat r \rangle_\alpha \pm \Delta_\alpha \hat r$ (and then the light blue cloud represents the quantum uncertainty).}
\end{figure}

Let $\alpha = re^{\imath \theta}$ be a complex representation of the local coordinates of the Schwarzschild slice, $z$ being a non-physical dimension needed for the embedding of the curved slice. Let $a = X+\imath  Y$ and $a^+=X-\imath Y$ be the creation and annihilation operators of a quantum harmonic oscillator, playing here the role of the noncommutative complex coordinate operators of the fuzzy version $\mathscr M$ of the Schwarzschild slice. $\mathscr F$, the associated Flock space, plays the role of $L^2(\mathscr M,\dnc X\otimes \dnc Y)$ (space of ``wave functions'' onto the fuzzy manifold). The third coordinate operator must satisfy a relation similar to eq.(\ref{SchwarzEq}), we set
\begin{equation}
  Z = f(\hat r)
\end{equation}
where $\hat r$ is the radius observable (the action of $f(\hat r)$ being defined by the functional calculus). We define this operator as being
\begin{equation}
  \hat r = \sqrt{a^+a} = \sum_{n=0}^{+\infty} \sqrt n |n\rangle \langle n|
\end{equation}
where $(|n\rangle)_{n \in \mathbb N}$ is the canonical basis of $\mathscr F$. We choose $\hat r$ with this expression and not as $\sqrt{X^2+Y^2} = \sqrt{a^+a+1/2}$ in order to have $\langle \alpha|\hat r^2|\alpha \rangle = |\alpha|^2$ (without the shift $1/2$) for $|\alpha \rangle$ a Perelomov coherent state of the harmonic oscillator (see appendix \ref{NCP}). We have then
\begin{equation}
  Z = \sum_{n=0}^{+\infty} f(\sqrt n) |n\rangle \langle n|
\end{equation}
Note that to apply this definition it needs to extend $f$ in the complex numbers to define its action onto negative values, the square root in the expression of $f$ being chosen in the Riemann sheet such that $\sqrt{-r} = -\imath \sqrt r$ (with $r>0$). $Z$ is then not self-adjoint, $Z^\dagger \not= Z$, and we set
\begin{equation}
  Z = \Re\, Z - \imath \Im\, Z
\end{equation}
with $\Re\, Z$ and $\Im\, Z$ the following positive self-adjoint operators:
\begin{eqnarray}
  \Re\, Z & = & 2\sqrt{r_S} \sum_{n=\lceil r_S^2 \rceil}^{+\infty} \sqrt{\sqrt n-r_S} |n\rangle \langle n| \\
  \Im\, Z & = & 2\sqrt{r_S} \sum_{n=0}^{\lfloor r_S^2 \rfloor} \sqrt{r_S-\sqrt n}|n\rangle \langle n|
\end{eqnarray}
In contrast with the classical slice, we cannot exclude the points $|\alpha|<r_S$ since $Z$ is defined in the $|n\rangle$-representation and not in a $|\alpha\rangle$-representation. A formula as $Z = \int_{|\alpha|>r_S} f(|\alpha|)|\alpha \rangle \langle \alpha | d\alpha d\bar \alpha$ makes no sense because the set $(|\alpha\rangle)_{\alpha \in \mathbb C}$ is overcomplete and not orthogonal ($|\langle \beta|\alpha\rangle|^2 = e^{-|\alpha-\beta|^2}$) \cite{Perelomov}.\\
$|\alpha \rangle$ represents a quantum state localized around $\alpha$, like a ``gaussian wave packet'' centered at $\alpha$ with half width at half maximum $\sqrt{\ln 2} \simeq 0.83$. The quantum event horizon is then not a precise circle on the slice, but is fuzzy with a Schwarzschild radius mean value $\langle r_S \rangle = \langle \alpha |\hat r|\alpha \rangle_{||\alpha|=r_S}$ and quantum uncertainty $\Delta r_S = \sqrt{\langle \alpha |\hat r^2|\alpha \rangle_{||\alpha|=r_S} - \langle \alpha |\hat r|\alpha \rangle_{||\alpha|=r_S}^2}$, as illustrated fig. \ref{rmoy}.
For small values of the parameter $r_S$, $\langle r_S \rangle$ is shifted from $r_S$; for example with $r_S=1.5$ we have $\langle r_S \rangle \pm \Delta r_S \simeq 1.36 \pm 0.61$. Since $r_S = 2M$ where $M$ is the black hole mass, this situation is similar to say that the singularity is delocalized following a Gaussian distribution and then that $M$ is smeared following a gaussian density. This is the idea used in \cite{Nicolini, Banerjee} to define a noncommutative geometry inspired black hole. In these papers, the authors compute the classical geometry induced by this smeared mass. In contrast, in this paper we rest at the quantum level by considering the quasi-coherent states of the fuzzy manifold $\mathscr M$. 

\subsection{Non-selfadjointness of the Dirac operator}
The Dirac operator for the fuzzy Schwarzschild slice $\mathscr M$ is
\begin{equation}\label{DiracOp}
  \slashed D_\alpha = \left(\begin{array}{cc} Z-z(\alpha) & a^+ - \bar \alpha \\ a- \alpha & -Z^\dagger + z(\alpha) \end{array} \right)
\end{equation}
where we restrict our attention onto the values $z$ for which $(\alpha,z(\alpha)) \in M_\Lambda$. $\slashed D_\alpha$ is not self-adjoint and we have chosen $Z \otimes |0\rangle \langle 0| - Z^\dagger \otimes |1\rangle \langle 1|$ and not $Z \otimes \sigma_z$ (where $(|0\rangle, |1\rangle)$ is the canonical basis of the spin state space $\mathbb C^2$) in order to the chosen Riemann sheet of $f$ be the same for all spin states of the fermionic string.\\
$-\imath \Im\, Z \otimes \id_2$ is a dissipator ($\Im\, Z$ is positive) involving that $||\psi(\tau)||$ decreases if $|\psi \rrangle$ is solution of the Schr\"odinger like equation $\imath |\dot \psi \rrangle = \slashed D_\alpha |\psi \rrangle$. In non-hermitian quantum mechanics \cite{Moiseyev}, dissipation is signature of the disappearance of the modeled quantum system. For example, in molecular physics, we can model a diatomic molecule as an anharmonic oscillator with Hilbert space $L^2([0,r_{max}],dr)$ where $r$ is the internuclear distance and $r_{max}$ is the end of the description box. After photodissociation, a wave packet representing the dissociated fragment spreads. To avoid unphysical reflections on $r_{max}$ an absorbing boundary dissipates this one before it reaches $r_{max}$. The dissipation represents then the disappearance of the molecule (when the wave packet reaches $r_{max}$ in the reality, we no longer have an associated molecule or an dissociated molecule but two independent atoms not modeled in the formalism), see \cite{Moiseyev}. We can interpret in a same manner the non-selfadjointness of $\slashed D_\alpha$. $\langle \alpha |\Im\, Z|\alpha \rangle \not\simeq 0$ if and only if $|\alpha|<r_S$. So the quantum dissipation occurs when the ``test particle'' is under the event horizon. In classical general relativity, for an external observer, the degrees of freedom of a particle under the event horizon disappear. The only one degree of freedom of a Schwarzschild black hole is its mass. Everything happens as if the particle was absorbed by the singularity. We have the same thing for our model of quantum black hole, except that due the quantum effects (quantum superpositions, fuzziness of the event horizon) this process is not instantaneous nor necessarily complete. $\|\psi(\tau)\|^2$ is then the survival probability of the test particle, or in other words, $1-\|\psi(\tau)\|^2$ is the probability of absorption of the particle by the black hole singularity. 

\subsection{Time observables}
The fermionic string state is solution of the Dirac equation
\begin{equation}\label{Diraceq}
  \imath |\dot \psi \rrangle = \slashed D_{\alpha(\tau)} |\psi(\tau) \rrangle
\end{equation}
for the transport of the probe D0-brane $\tau \mapsto \alpha(\tau)$, $\tau$ being the clock of an observer comoving with the probe brane. We denote by $t$ the clock of an observer at ``infinity'' comoving with the black hole (the Schwarzschild time coordinate), and by $t(\tau)$ the time measured by this clock when the observer watches the clock ``attached'' to the probe D0-brane. In contrast with space coordinates $X$ and $Y$, the coordinate time $t$ is not quantized in the model. We can introduce a kind of time quantization with the following construction. Let $t'(\tau)$ be the time measured with the clock of the observer at infinity when this one watches a clock ``attached'' to the other end of the fermionic string. At this stage, we suppose that this observable is classical and that the ``proper time'' of this end is the same than the one of the probe D0-brane (even so the state of the other end is not classical). We can then write than $\psi$ depends on $\tau$ directly from the D0-brane and from the other end: $\psi(\tau,t'(\tau))$, and so:
\begin{equation}
  \frac{d|\psi\rrangle}{d\tau} = \frac{\partial |\psi\rrangle}{\partial \tau} + \frac{\partial |\psi\rrangle}{\partial t'} \dot t'
\end{equation}
\begin{equation}
  \Rightarrow \imath \frac{\partial |\psi\rrangle}{\partial \tau} = \left(\slashed D_{\alpha(\tau)} - \imath \dot t' \frac{\partial}{\partial t'} \right) |\psi \rrangle
\end{equation}
We make the hypothesis than the movement of the other end is geodesic, and then that
\begin{equation}
  \dot t' = \frac{\epsilon'}{1-\frac{r_S}{r}}
\end{equation}
where $\epsilon'$ is the constant of motion of a Schwarzschild geodesic (total energy by mass unit). Now we want to modify the model to have a quantized time for the other end which is a quantum system. We extend the Hilbert space as $\mathscr H^+ = \mathbb C^2 \otimes \mathscr F \otimes L^2(\mathbb R,dt')$, where $\mathscr F \sim_{NC} L^2(\mathscr M,\dnc X \otimes \dnc Y)$ is the space of space wave functions, $L^2(\mathbb R,dt')$ is the space of time wave functions and we introduce the operator
\begin{equation}
  \hat{\dot t} = \dot t'(\hat r) = \epsilon' \sum_{n=1}^{\infty} \frac{1}{1-\frac{r_S}{\sqrt n}} |n\rangle \langle n|
\end{equation}
the Dirac equation in the extended Hilbert space being
\begin{equation}
  \imath \frac{\partial |\psi\rrrangle}{\partial \tau} = \left(\slashed D_{\alpha(\tau)} - \imath \hat{\dot t} \otimes \frac{\partial}{\partial t'} \right) |\psi \rrrangle
\end{equation}
The replacement of $\dot t'$ by $\hat {\dot t}$ takes into account that the other end of the fermionic string does not follow a classical trajectory but is a quantum ``wave packet'' smeared in the embedding space (because of the fuzziness of the quantum Schwarzschild slice $\mathscr M$ on which this end is attached). $E = -\imath \hat{\dot t} \otimes \partial_{t'}$ plays in this toy model the role of an energy observable. Since $e^{\imath \omega t'}$ is eigenvector of $-\imath \partial_{t'}$ with eigenvalue $\omega$, we can choose without loss of generality $|\psi(\tau) \rrrangle = |\psi(\tau) \rrangle \otimes |e^{\imath \omega t'} \rangle$ and then
\begin{equation}\label{DiracEeq}
  \imath \frac{\partial |\psi \rrangle}{\partial \tau} = \left(\slashed D_{\alpha(\tau)} + \omega \hat{\dot t}\right)|\psi \rrangle
\end{equation}
We can view $\omega$ as a mass parameter, and then the true parameter of the extended model $\omega \epsilon'$ is the total energy of the end of the string attached on $\mathscr M$. Note that if $r_S^2 \in \mathbb N$, $\hat{\dot t}$ is singular with an infinite eigenvalue for $|n=\sqrt{r_S}\rangle$. This implies that $\langle n=\sqrt{r_S}|\psi \rrangle = 0$ ($\forall \tau$) ($|n=\sqrt{r_S}\rangle$ becomes a forbidden state). In order to avoid the mathematical and numerical difficulties due to the singular value, we suppose that $r_S^2 \not\in \mathbb N$ in the rest of this paper.\\
By construction the time dilation observable $\hat {\dot t}$ plays the role of a barrier in the neighborhood of the event horizon. In classical general relativity, the time measured by the clock of the observer at infinity is dilated with respect to the proper time when the observed particle approaches the event horizon (and becomes infinite at the even horizon). The observer never sees the particle cross the event horizon. This effect is reproduced for the observation of the ``wave packet'' of the end of the string attached to $\mathscr M$ by the barrier of $\hat{\dot t}$.\\

The model includes the time dilation operator $\hat{\dot t}$ and not directly the observable of the quantized time coordinate. Let $\langle t' \rangle$ be the mean value of the time measured by the observer at infinity when this one watches a clock ``attached'' to the end of the string on $\mathscr M$ (and so when the observer compares its clock with the quantized time). Heuristically we can set that
\begin{equation}
  \langle t' \rangle = \int_0^\tau \frac{\llangle \psi|\hat{\dot t}|\psi \rrangle}{\|\psi\|^2} d\tau
\end{equation}
Indeed, if it had existed, an observable $\hat t'$ would have satisfied $\frac{d\lllangle \hat t' \rrrangle}{d\tau} = \imath \lllangle [\slashed D_\alpha+E,\hat t']\rrrangle$. If we substitute $t'$ (as the operator multiplying by the quantum variable $t'$) to $\hat t'$, we have $[\slashed D_\alpha+E,\hat t'] \leadsto [\slashed D_\alpha+E,t'] = [E,t'] = \hat{\dot t}$. In the same way we can set the following estimation
\begin{equation}
  \Delta {t'}^2 \sim \int_0^\tau \Delta \hat{\dot t}^2 d\tau = \int_0^\tau (\llangle \hat{\dot t}^2 \rrangle - \llangle \hat{\dot t} \rrangle^2) d\tau
\end{equation}

\section{Schwarzschild quasi-coherent state}
We search the quasi-coherent states of the fuzzy Schwarzschild slice $\mathscr M$ (defined by $(a,a^+,Z$)) and the associated eigenmanifold $M_\Lambda$ (classical manifold closest to the $\mathscr M$ in the sense that its points are the mean values of the coordinate observables of $\mathscr M$ with minimal quantum dispersion). Since $\slashed D_\alpha$ is not self-adjoint, the associated eigenvalue $\lambda_0(\alpha)$ is not zero, but is purely imaginary:
\begin{equation}\label{QCSeq}
  \slashed D_\alpha |\Lambda(\alpha) \rrangle = \lambda_0(\alpha)|\Lambda(\alpha) \rrangle, \quad \Re\, \lambda_0 = 0
\end{equation}
This definition is consistent with the interpretation of non-hermitian quantum mechanics \cite{Moiseyev}. $\slashed D_\alpha$ plays the role of an Hamiltonian in eq.(\ref{Diraceq}) which has the structure of a Schr\"odinger equation. The interpretation of the complex number $\lambda_0 \in \Sp(\slashed D_\alpha)$ is the following:
\begin{itemize}
\item $\Re\, \lambda_0$ is the ``energy'' measured by $\slashed D_\alpha$ when the system is in the state $|\Lambda \rrangle$ (in fact it is the square root of the displacement energy -- see \cite{Schneiderbauer} -- the displacement energy observable being $\slashed D_\alpha^2$ in the self-adjoint cases);
\item $-\Im\, \lambda_0$ is the inverse of the characteristic time of dissipation in the state $|\Lambda \rrangle$.
\end{itemize}
We can see this by considering the solution of eq.(\ref{Diraceq}) when $\alpha$ is frozen and with $|\psi(0)\rrangle = |\Lambda(\alpha)\rrangle$:
\begin{equation}
  |\psi(\tau) \rrangle = e^{-\imath \Re\, \lambda_0(\alpha) \tau} e^{\Im\, \lambda_0(\alpha) \tau} |\Lambda(\alpha) \rrangle
\end{equation}
$|\psi \rrangle$ has an oscillating phase of frequency $\Re\, \lambda_0(\alpha)$ (which will generate Rabbi oscillations if we choose $|\psi(0)\rrangle$ has a superposition of two different eigenvectors). The norm exponentially decreases as $\|\psi(\tau)\|^2 = e^{2\Im\, \lambda_0(\alpha) \tau}$ ($\Im \lambda_0<0$). The survival time of the particle is then $\tau_{surv}(\alpha) = \frac{-1}{\Im\, \lambda_0(\alpha)}$ (after a duration $\tau_{surv}$ the probability for which the particle be absorbed by the singularity is larger than 85\%). The quasi-coherent states being defined as being the states minimizing the displacement energy, we must have $\Re\, \lambda_0=0$. But $\Im\, \lambda_0 \not=0$ because of the process of absorption by the black hole singularity. Nevertheless, we expect that $\Im\, \lambda_0(\alpha) \simeq 0$ for $|\alpha| > r_S$.\\

The eigenmanifold is then defined as
\begin{equation}
  M_\Lambda = \{(\alpha,z(\alpha)) \in \mathbb C \times \mathbb R, \text{ s.t. } \Sp(\slashed D_\alpha)\cap \imath \mathbb R \not= \varnothing \}
\end{equation}
The solving of eq.(\ref{QCSeq}) needs to find the three quantities: $|\Lambda(\alpha)\rrangle \in \mathscr H$, $\lambda_0(\alpha) \in \imath \mathbb R$ and $z(\alpha) \in \mathbb  R$. Appendix \ref{nummeth} presents the numerical method used to this.

\subsection{Emergent geometry}
\begin{figure}
  \includegraphics[width=8cm]{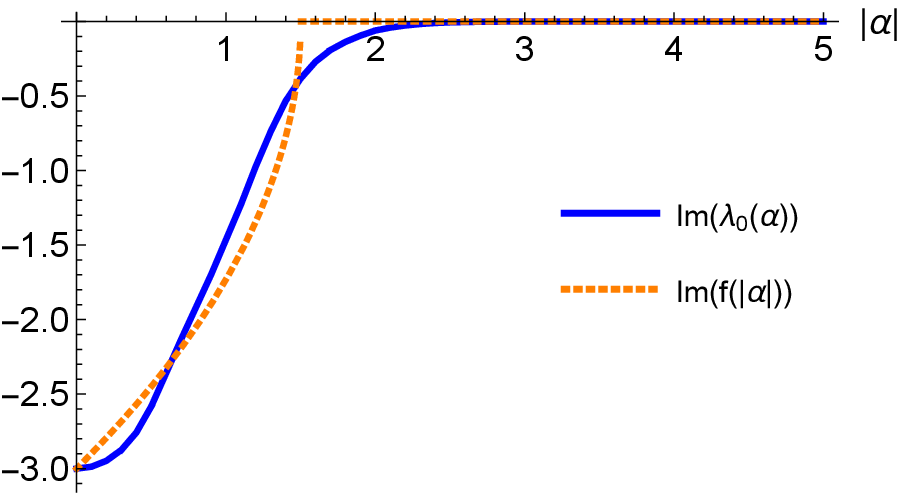}\\
  \includegraphics[width=8cm]{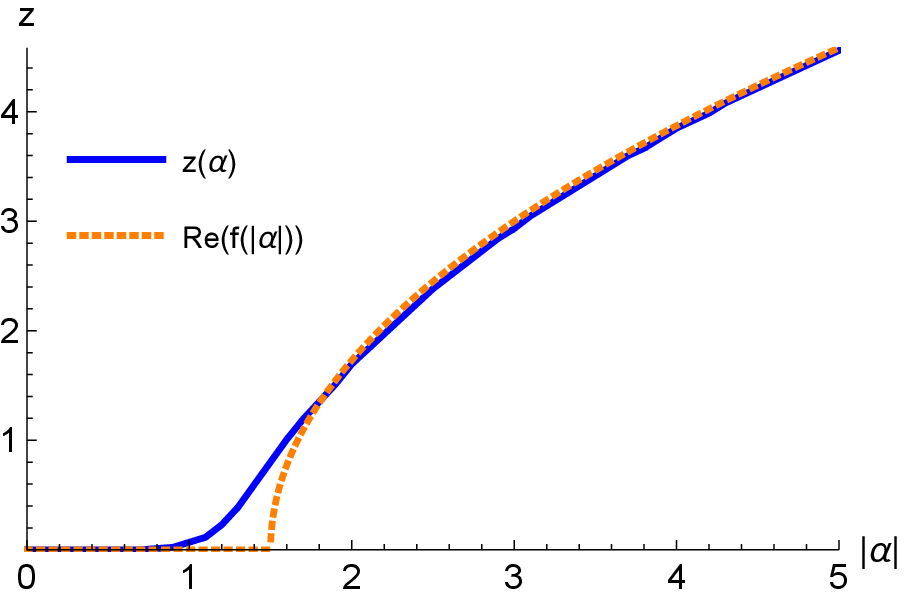}\\
  \caption{\label{NCeigenval} Up: $\Im\, \lambda_0(\alpha)$ (inverse of the dissipation time) compared to $\Im\, f(|\alpha|)$. Down: $z(\alpha)$ compared to $\Re\, f(|\alpha|)$. $\Im\, \lambda_0$ and $z$ are solutions of eq.(\ref{QCSeq}) with $r_S=1.5$.}
\end{figure}

\begin{figure}
  \includegraphics[width=8cm]{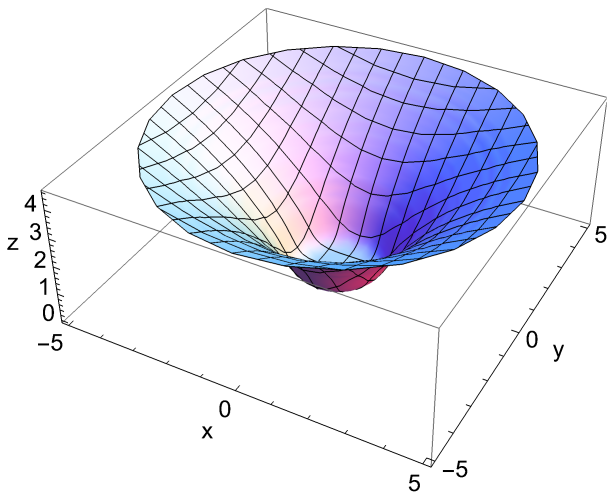}\\
  \caption{\label{MLambda} Eigenmanifold $M_\Lambda$ of the fuzzy Schwarzschild slice $\mathscr M$ ($M_\Lambda$ is the set of the mean values of $\mathscr M$ in the quasi-coherent state), with $r_S=1.5$.}
\end{figure}

\begin{figure}
  \includegraphics[width=8cm]{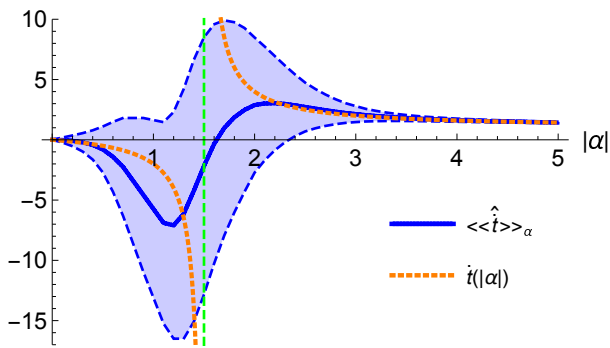}\\
  \caption{\label{dotT} $\llangle \Lambda(\alpha)|\hat{\dot t}|\Lambda(\alpha)\rrangle$ compared to the classical quantity $\dot t(|\alpha|) = \frac{\epsilon'}{1-\frac{r_S}{|\alpha|}}$. The dashed lines correspond to $\llangle \hat{\dot t} \rrangle_\alpha \pm \Delta_\alpha \hat{\dot t}$ (the light blue cloud represents the quantum uncertainty). The parameters are $r_S=1.5$ and $\epsilon'=1$.}
\end{figure}

The functions $\alpha \mapsto \Im\, \lambda_0(\alpha)$ and $\alpha \mapsto z(\alpha)$ are drawn fig.(\ref{NCeigenval}).
As expected $\Im\, \lambda_0(\alpha) \simeq 0$ for $\alpha > r_S$ and roughly follows the imaginary part of $f(|\alpha|)$. As expected $z(\alpha)$ presents the profile of the classical Schwarzschild slice except that the decreasing in the neighborhood of $r_S$ is softened (in accordance with the fuzziness of the quantum event horizon $\langle r_S \rangle \pm \Delta r_S \simeq 1.36 \pm 0.61$ when the classical parameter is $r_S=1.5$). The embedding of $M_\Lambda$ in $\mathbb R^3$ is represented fig.\ref{MLambda}.
For the aspects concerning the time, we have drawn the mean value of $\hat {\dot t}$ in the quasi-coherent state fig. \ref{dotT}.
In $\alpha$-representation $\hat {\dot t}$ plays then well the role of an potential barrier in front of the event horizon (a small barrier in average, more important by taking into account the quantum uncertainty), preventing the crossing of the event horizon by the particle observed by an observer at infinity comoving with the quantum black hole ($\hat{\dot t}$ is the observable of time dilation governing the quantum end of the string attached on $\mathscr M$ measured by the observer's clock). We see also a potential well behind the event horizon trapping the particle inside the black hole.

\subsection{Structure of the quasi-coherent state}
\begin{figure}
  \includegraphics[width=8cm]{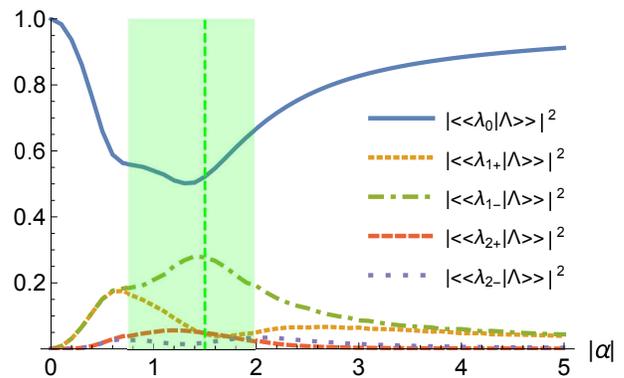}\\
  \caption{\label{QCSdecomp} Main probabilities of occupation of the states $|\lambda_{n\pm}(\alpha) \rrangle$ when the system is in the quasi-coherent state with respect to $|\alpha|$. The vertical dashed line represents the classical event horizon $r_S=1.5$, the light green cloud represents the fuzzy event horizon ($\langle r_S\rangle \pm \Delta r_S$).}
\end{figure}

\begin{figure}
  \includegraphics[width=8cm]{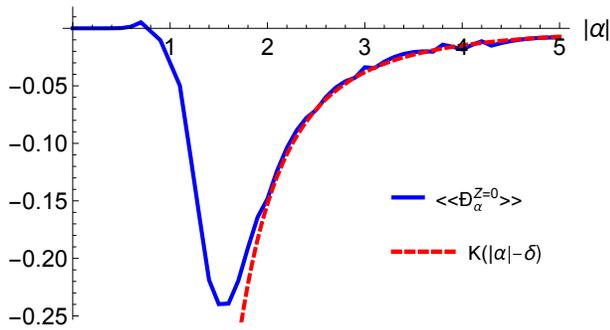}\\
  \caption{\label{deform} Average deformation $\llangle \Lambda(\alpha)|\slashed D_\alpha^{Z=0}|\Lambda(\alpha)\rrangle$ with respect to the plane case in the quasi-coherent state of the fuzzy Schwarzschild slice, compared to the shifted Gauss curvature $K(r-\delta) = - \frac{r_S}{2(r-\delta)^3}$ of the classical Schwarzschild slice. The shift $\delta$ is introduced to take into account the fuzziness of the singularity and the smoothness of the average geometry (see fig.\ref{NCeigenval}). For $r_S > 1$, numerical tests show that a good value of the shift is $\delta = 0.6(r_S-1)$ (here with $r_S=1.5$).}
\end{figure}

Let $(|n_\alpha\rangle)_{n \in \mathbb N}$ be the (local) orthonormal basis of $\mathscr F$ defined by
\begin{equation}
  |n\rangle_\alpha = \frac{(a^+-\bar \alpha)^n}{\sqrt{n!}} |\alpha \rangle
\end{equation}
$|\alpha \rangle$ being the Perelomov coherent state of the harmonic oscillator algebra. This basis is ``local'' in the sense that it is defined for a point $\alpha$ and is associated with an harmonic oscillator re-centred on $\alpha$ (we can interpret $|n\rangle_\alpha$ as a quantum state with a local excitation of amplitude $n$ at $\alpha$). Let $(|\lambda_{n\pm}(\alpha) \rrangle$ be the local basis of $\mathscr H = \mathbb C^2 \otimes \mathscr F$ defined by
\begin{eqnarray}
  |\lambda_0(\alpha) \rrangle & = & |0\rangle \otimes |0\rangle_\alpha \\
  |\lambda_{n\pm}(\alpha) \rrangle & = & \frac{1}{\sqrt 2}(|0\rangle \otimes |n\rangle_\alpha \pm |1\rangle \otimes |n-1\rangle_\alpha)
\end{eqnarray}
This is the eigenbasis of the Dirac operator in the case of the noncommutative plane ($Z=0$) (see appendix \ref{NCP}). $|\lambda_0\rrangle$ is the quasi-coherent state of the noncommutative plane whereas $|\lambda_{n\pm} \rrangle$ (with $n>0$) are states for which the displacement energy is $n$. Since our model of fuzzy Schwarzschild slice is a deformation of the noncommutative plane, this basis is pertinent to study the quasi-coherent state. We expect in particular that $|\Lambda(\alpha)\rrangle \sim |0\rangle \otimes |\alpha \rangle = |\lambda_0(\alpha) \rrangle$ for $|\alpha| \to +\infty$ (the classical Schwarzschild geometry being asymptotically flat). The decomposition of $|\Lambda(\alpha)\rrangle$ onto the local basis is represented, fig. \ref{QCSdecomp}.
We see that $|\Lambda(\alpha)\rrangle$ significantly differs from the quasi-coherent state of a flat noncommutative plane $|\lambda_0(\alpha) \rrangle$ essentially in the neighborhood of the fuzzy event horizon and inner the quantum black hole. By construction $|\llangle \lambda_0(\alpha)|\Lambda(\alpha) \rrangle|^2$ is the probability of flatness (probability for which an observer comoving with the D0-brane at $\alpha$ finds the space flat by measures on the test particle -- prepared in the quasi-coherent state $|\Lambda \rrangle$ --). The excited states $\{|\lambda_{n\pm}(\alpha)\rrangle\}_{n>0}$ can be viewed as states of local (positive or negative) deformations of the plane with magnitude $\sqrt n$ at $\alpha$ (since from the viewpoint of the fermionic string, they correspond to a probe D0-brane far from $\mathscr M$ at the mean distance $\sqrt n$ and with a dispersion $\Delta_{\lambda_{n\pm}} X \Delta_{\lambda_{n\pm}} Y = n > \frac{1}{4} = \Delta_{\lambda_0} X \Delta_{\lambda_0} Y$). $|\llangle \lambda_{n\pm}(\alpha)|\Lambda(\alpha)\rrangle|^2$ is then the probability to find a deformation with respect of the plane of magnitude $\sqrt n$ at $\alpha$ (by measurement on the test particle by an observer comoving with the probe D0-brane at $\alpha$). To confirm this, consider the mean value of the Dirac operator of the noncommutative plane $\slashed D_\alpha^{Z=0} = {\scriptstyle \left(\begin{array}{cc} 0 & a^+-\bar \alpha \\ a-\alpha & 0 \end{array}\right)}$ in the quasi-coherent state of the fuzzy Schwarzschild slice:
\begin{equation}
  \llangle \Lambda(\alpha)|\slashed D_\alpha^{Z=0}|\Lambda(\alpha)\rrangle = \sum_{{n\in \mathbb N^*} \atop {\varsigma\in\{\pm\}}} |\llangle \lambda_{n\varsigma}(\alpha)|\Lambda(\alpha)\rrangle|^2 \varsigma \sqrt n
\end{equation}
This average deformation is compared with the Gauss curvature of the classical Schwarzschild slice fig.\ref{deform}.
$\llangle \slashed D_\alpha^{Z=0} \rrangle$ follows then a law similar to the classical Gauss curvature (for $|\alpha|>r_S$) confirming that it can be then interpreted as the average deformation with respect to the plane case, and that $\{|\lambda_{n\pm}\rrangle\}_{n>0}$ can be interpreted as states of deformation. The bounded character of the average deformation and its fall behind the event horizon is in accordance with the smoothness of the quasi-coherent geometry with respect to the classical geometry (see fig.\ref{NCeigenval}).\\

\begin{figure}
  \includegraphics[width=8cm]{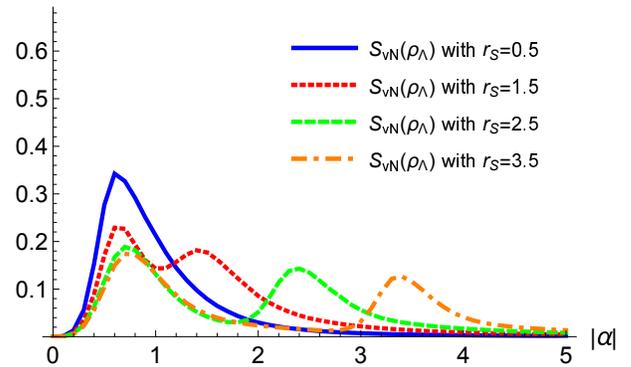}
  \caption{\label{entropyVSrs} von Neumann entropy of the spin state of the fermionic string in the quasi-coherent state for different values of the classical Schwarzschild radius $r_S$.}
\end{figure}

It is interesting to study the behavior of the spin of the fermionic string, because in classical general relativity, the transport of an angular momentum in a spacetime can induce a rotation of this one (as for example by Thomas or de Sitter precessions). But in the quantum context, the effects of a quantum spacetime onto a quantum spin can also induced quantum entanglement. The state of the string's spin is described by the density matrix
\begin{equation}
  \rho_\Lambda(\alpha) = \tr_{\mathscr F}|\Lambda(\alpha)\rrangle \llangle \Lambda(\alpha) |
\end{equation}
where $\tr_{\mathscr F}$ denotes the partial trace over the states of $\mathscr F$. In particular, we can consider the von Neumann entropy $S_{vN}(\rho_\Lambda) = - \tr(\rho_\Lambda \ln \rho_\Lambda)$ fig.(\ref{entropyVSrs}).
The quasi-coherent state is entangled in the neighborhood of the fuzzy event horizon and also in a region around $|\alpha|\simeq 0.7$ independently of $r_S$. The interpretation of this second peak of entropy is not clear, but due to its position (which is independent of the black hole parameter), we can think that it is related to the border of the fuzzy singularity. 

\section{Quantum dynamics}
Now we want study the dynamics governed by eq.(\ref{DiracEeq}) with $|\psi(0)\rrangle = |\Lambda(\alpha(0))\rrangle$ for different worldlines $\tau \mapsto (t(\tau),\alpha(\tau))$ of the probe D0-brane. We control the probe D0-brane to drive the fermionic string (if the state remains a quasi-coherent state during the whole of the dynamics, the probe D0-brane drives the point-like particle defined by a gaussian wave packet with minimal width in the $\alpha$-representation). $\frac{d\slashed D_\alpha}{d\tau} = - \dot \alpha \sigma_- -\dot {\bar \alpha} \sigma_+ - (\frac{\partial z}{\partial \alpha} \dot \alpha+\frac{\partial z}{\partial \bar \alpha} \dot{\bar \alpha}) \sigma_z$, the regime of the dynamics is then essentially governed by $|\dot \alpha|$. In the case of the transport of a D0-brane in a noncommutative plane (appendix \ref{NCP}), we have two extremal regimes: an adiabatic regime for small velocities of the probe D0-brane where the state remains an instantaneous quasi-coherent state; and a sudden regime for large velocities. The interest of the sudden regime is low, the probe D0-brane being transported at a velocity close to the speed of light, the ``wave packet'' of the other end of the string in the $\alpha$-representation cannot follow it and remains at its initial position. In the case of the fuzzy Schwarzschild spacetime, we expect similar behaviors with respect to the magnitude of $|\dot \alpha|$, but due to the irregularity of the event horizon in the Schwarzschild coordinates, the dynamical behavior will be strongly affected in the neighborhood of the fuzzy event horizon.

\subsection{Decoherence processes}
Three quantities must be particularly monitored during the dynamics:
\begin{itemize}
\item $\|\psi\|^2$ the survival probability of the fermion.
\item The fidelity to a quasi-coherent state:
\begin{equation}
  F(\Lambda|\psi) = \sup_\alpha \frac{|\llangle \Lambda(\alpha)|\psi \rrangle|^2}{\|\psi\|^2}
\end{equation}
which is equal to 1 if $|\psi \rrangle$ is proportional to a quasi-coherent state for some $\alpha$ and 0 if $|\psi \rrangle$ is orthogonal to the quasi-coherent states for all $\alpha$. $F(\Lambda|\psi)$ measures then the sustain of $|\psi \rrangle$ as a quasi-coherent state corresponding to a classical point particle (gaussian wave packet in $\alpha$-representation with minimal Heisenberg uncertainties). We can then consider $F(\Lambda|\psi)$ as a measure of the spatial coherence. Moreover, since in adiabatic regimes $|\psi(\tau)\rrangle \propto |\Lambda(\alpha(\tau))\rrangle$, $F(\Lambda|\psi(\tau))$ is also a measure of the adiabaticity of the dynamics.
\item The purity:
  \begin{equation}
    \mathcal P(\rho) = \tr(\rho^2) \quad \text{with } \rho = \tr_{\mathscr F}\frac{|\psi \rrangle \llangle \psi|}{\|\psi\|^2}
  \end{equation}
  which is equal to 1 if $|\psi \rrangle$ is separable and 0 if $|\psi \rrangle$ is maximally entangled. The purity is then a measure of the entanglement between the spin degree of freedom and the space degree of freedom described by $\mathscr F$. In classical general relativity, the transport of an angular momentum in a curved spacetime involves rotation of this one (with Thomas and de Sitter precession phenomena for example). In quantum gravity, we can expect that the transport of a spin in a fuzzy spacetime involves not only a precession but also a loss of purity: the spin state becomes a statistical mixture of spin pure states.
\end{itemize}
We can see the decreasing of one of these three quantities as a decoherence process: a fall of the survival probability characterizes the disappearance of the fermion by absorption by the black hole singularity; a fall of the fidelity to a quasi-coherent state characterizes the decreasing of the spatial coherence (the state can no longer be correctly described by a gaussian wave packet in the $\alpha$-representation), a fall of the purity characterizes the increasing of the entanglement between the spin and the quantum spacetime.

\subsection{Fall on a quantum black hole}
\begin{figure}
  \includegraphics[width=8cm]{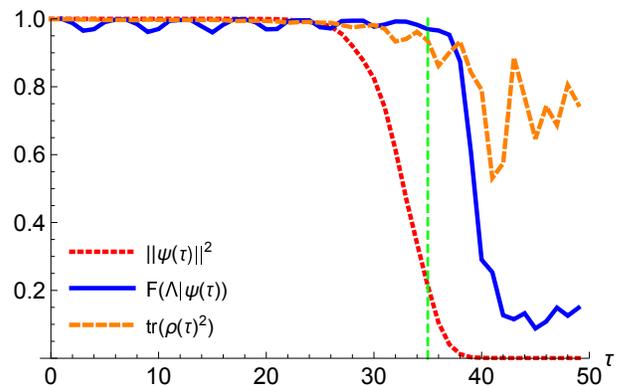}\\
  \caption{\label{coheUni} Evolutions of the survival probability, of the spatial coherence and of the purity of the fermionic state for a radial transport of the probe D0-brane to the singularity with $|\dot r|=u=0.1$ (corresponding to an initial speed $v(0) \simeq 0.56$). $r_S = 1.5$ (the vertical line corresponds to the passage of the probe D0-brane by the classical event horizon) and $\omega \epsilon'=0$.}
\end{figure}

\begin{figure}
  \includegraphics[width=8cm]{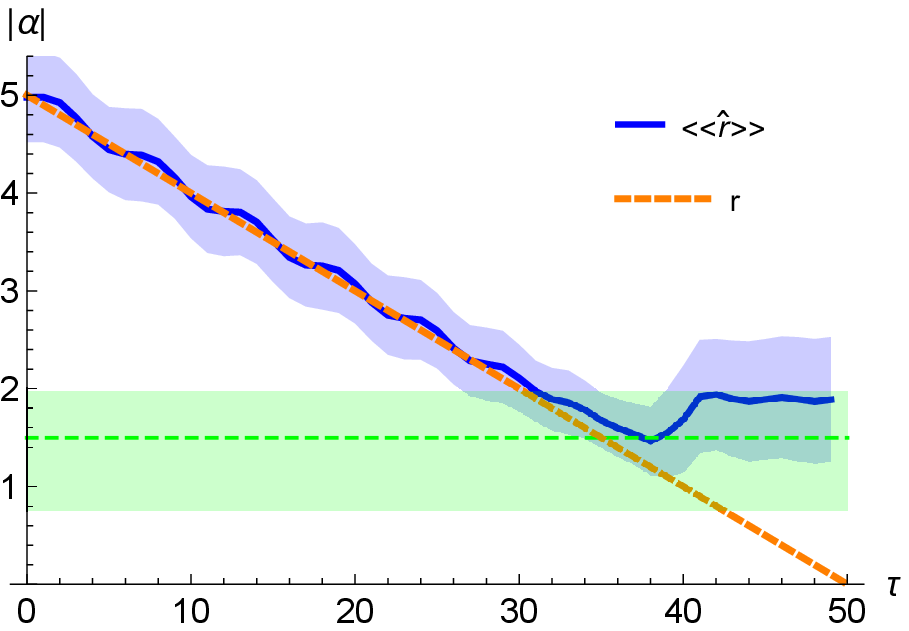}\\
  \includegraphics[width=8cm]{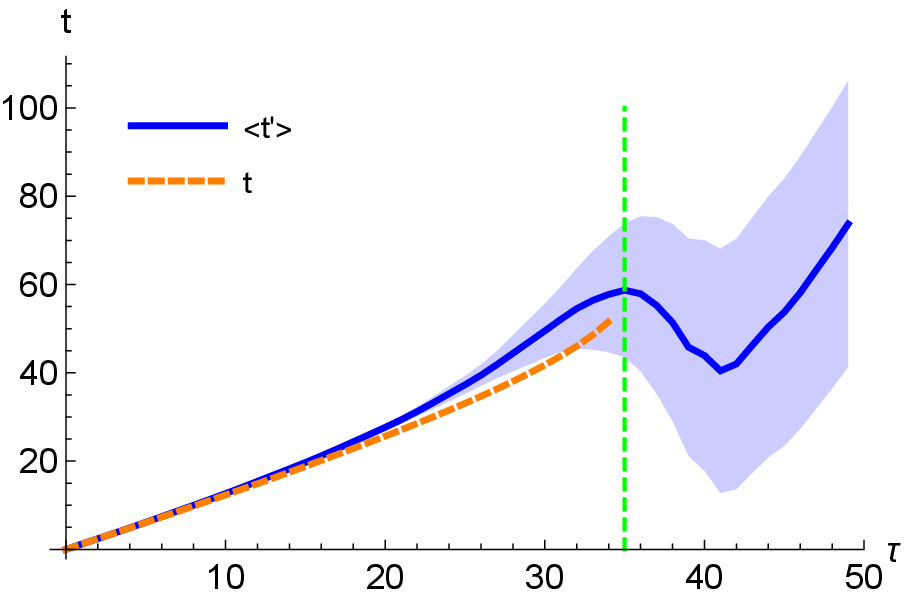}\\
  \caption{\label{rtmoychuteuni} $\llangle \hat r \rrangle = \frac{\llangle \psi|\hat r|\psi \rrangle}{\|\psi\|^2}$ (mean value of the radial coordinate of the end of the string attached on $\mathscr M$) and $\langle t'\rangle = \int_0^\tau \frac{\llangle \psi|\hat{\dot t}|\psi \rrangle}{\|\psi\|^2} d\tau$ (mean value of the time of the end of the string measured by the clock of an observer at infinity comoving with the black hole) compared to the radial and time coordinates of the probe D0-brane (for $|\dot r|=u=0.1$). The light blue clouds correspond to the quantum uncertainties. The parameters are $r_S=1.5$ and $\omega \epsilon'=0$. Up, the dashed horizontal line is the classical event horizon (the light green cloud corresponds to the fuzzy event horizon $\langle r_S \rangle \pm \Delta r_S$); down, the dashed vertical line indicates the time where the probe D0-brane crosses the classical event horizon.}
\end{figure}

\begin{figure}
  \includegraphics[width=8cm]{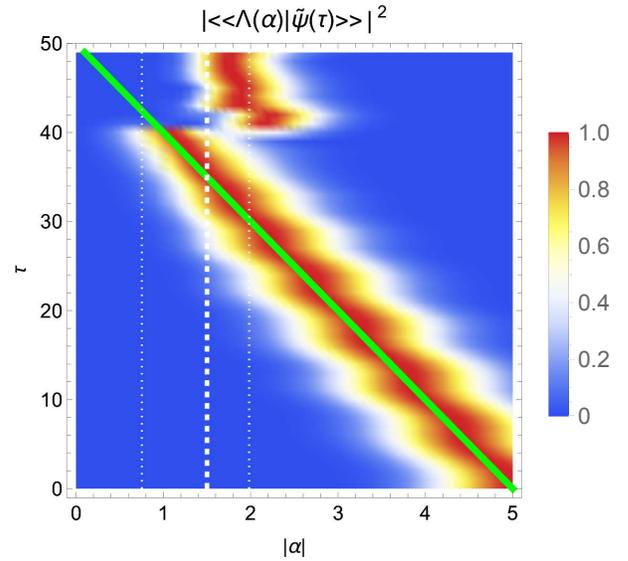}
  \caption{\label{WPuni} Wave packet $\psi(\alpha,\tau)$ representing the spatially coherent part of the end of the string state compared to the radial coordinate of the probe D0-brane (green path) when $|\dot r|=u=0.1$. The white dashed lines represent the classical event horizon $r_S=1.5$ and the fuzzy event horizon $\langle r_S\rangle \pm \Delta r_S$. $\omega \epsilon'=0$.}
\end{figure}

We consider first an uniform radial transport of the probe D0-brane to the singularity with $\dot r = -u$ (with $u>0$ constant), corresponding to $\dot t = \sqrt{(1-\frac{r_S}{r})^{-1}+(1-\frac{r_S}{r})^{-2} u^2} \equiv \frac{1}{\sqrt{1-v^2}}$ (with $v$ the velocity of the D0-brane). For a small value of $u$, the evolutions of the coherences are drawn fig.\ref{coheUni}.
As could be expected, strong decoherences occur at the event horizon. In particular, the spatial decoherence falls (the speed at the neighborhood of $r_S$ becomes equal to the speed of light, involving a strong non-adiabaticity) and the survival probability falls (inside the black hole the fermion is absorbed by the singularity). The mean values of $\hat r$ and $t'$ are drawn fig.\ref{rtmoychuteuni}.
The other end of the string follows the probe D0-brane, but it seems to bounce on the event horizon and to stagnate in its neighborhood. This can be confirmed by examining the wave packet of the $\alpha$-representation of the end of the string:
\begin{equation}
  \psi(\alpha,\tau) = \llangle \Lambda(\alpha)|\tilde \psi(\tau) \rrangle
\end{equation}
where $|\tilde \psi(\tau) \rrangle = \frac{|\psi(\tau) \rrangle}{\sup_\beta \llangle \Lambda(\beta)|\psi(\tau) \rrangle}$ is the renormalized state in order to normalize its spatial coherence at 1 at each time (this permits to examine the evolution of the spatially coherent part of $|\psi \rrangle$). $\psi(\alpha,\tau)$ is represented fig.(\ref{WPuni}).
We have considered the case $\omega \epsilon'=0$, so the particle is not submitted to the barrier generated by $\hat{\dot t}$. We can interpret the behavior at the event horizon as being similar to the reflection of a wave packet by a rough potential well. A part of the wave packet is reflected by the black hole (which can be viewed as a potential well in $\alpha$-representation fig.\ref{NCeigenval}). But driven by the probe D0-brane, this wave packet part returns on event horizon and bounces again and again (fig.\ref{WPuni}). But note that in fig.\ref{rtmoychuteuni}\&\ref{WPuni} the state $|\psi \rrangle$ is renormalized. In fact, as we can see it fig.\ref{coheUni}, the most of the state is transmitted by the event horizon and is quickly dissipated (absorbed by the singularity). So the mean position in fig.\ref{rtmoychuteuni}\&\ref{WPuni} after the arrival at the event horizon corresponds to the position of the tiny part which survives by reflection. The transmitted part does not contribute to the mean position since this one disappears quickly.\\

\begin{figure}
  \includegraphics[width=4cm]{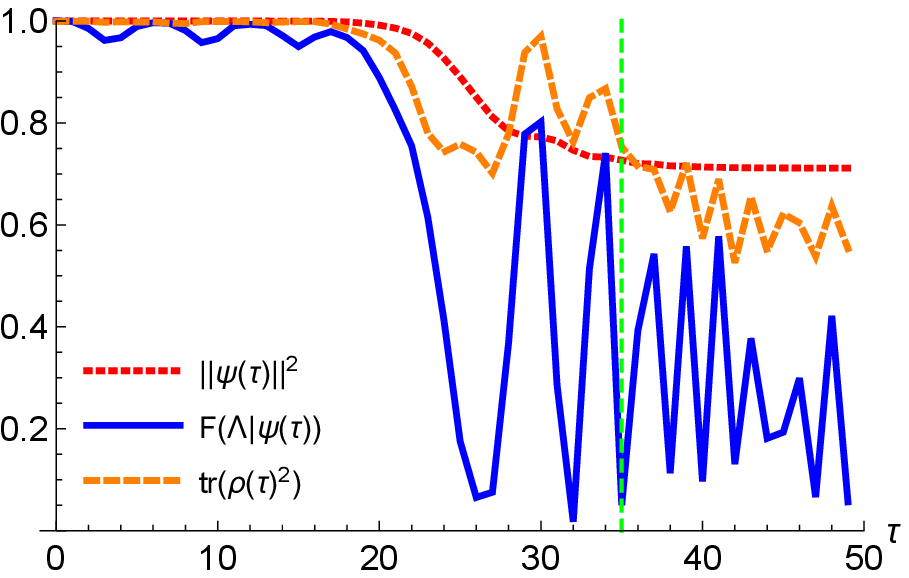}\includegraphics[width=4cm]{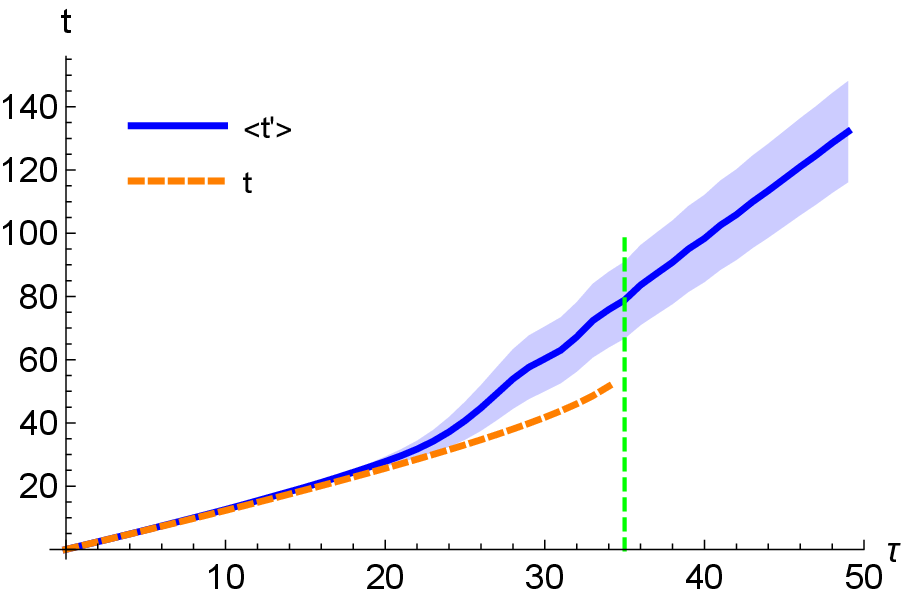}\\
  \includegraphics[width=4cm]{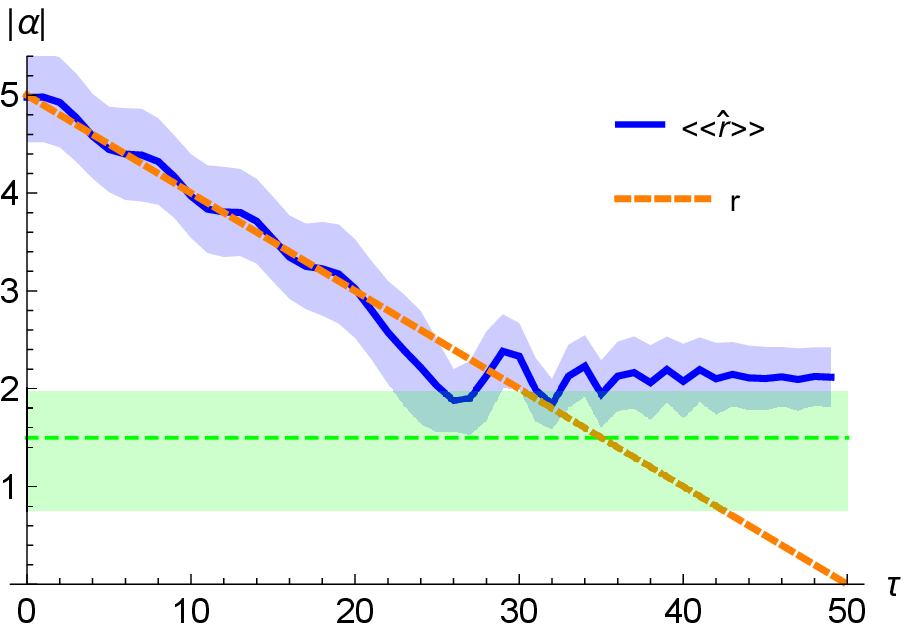}\includegraphics[width=4cm]{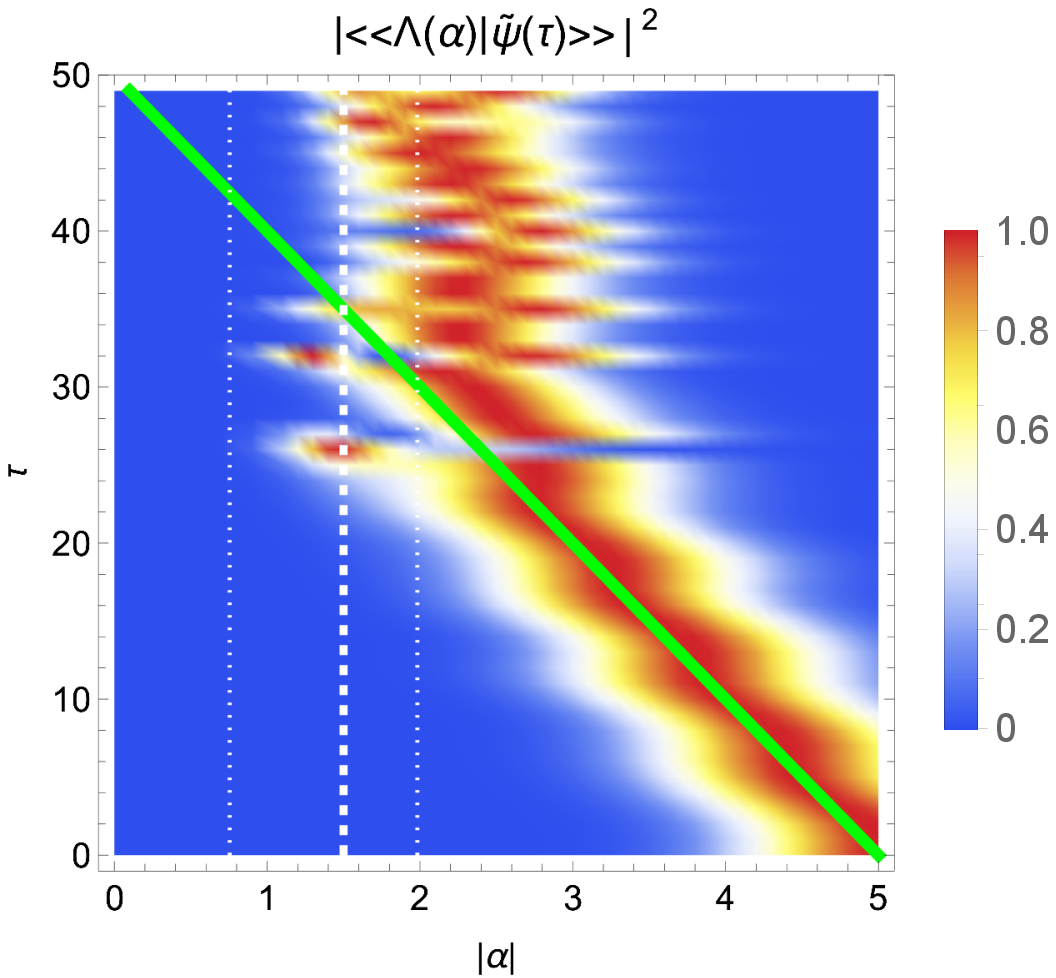}\\
  \caption{\label{uni2} Same as fig.\ref{coheUni} (up left), fig.\ref{rtmoychuteuni} (up right and down left) and fig.\ref{WPuni} (down right), but with $\omega \epsilon'=0.5$.}
\end{figure}

The situation differs if we consider the case where $\omega \epsilon' \gg 0$, fig.\ref{uni2}.
The barrier on the event horizon induced by $\omega \hat{\dot t}$ (fig.\ref{dotT}) reflects a large part of the wave packet, inducing that the dissipation (which concerns the transmitted part) is not complete. The rebounds on the event horizon (now induced by the barrier) concerns now a particle with a non-negligible survival probability. These ones imply strong oscillations of the spatial coherences. It can be counterintuitive that a quantum black hole has more difficulties for absorbing a heavy particle $\omega \epsilon'\gg 0$ than a light one $\omega \epsilon' \simeq 0$. But note that black hole mass is $M = \frac{r_S}{2}$. In the example fig.\ref{uni2} where the particle essentially bounces on the event horizon, we have $M=0.75$ and $\omega \epsilon'=0.5$. It is not surprising that the quantum black hole has difficulties to absorb a particle almost as big as it. With $\omega \epsilon'=1>M$, the survival probability rests equal to 1 during the whole dynamics, the reflection by the event horizon is then total. The quantum black hole cannot absorb a particle bigger than itself.\\

\begin{figure}
  \includegraphics[width=4cm]{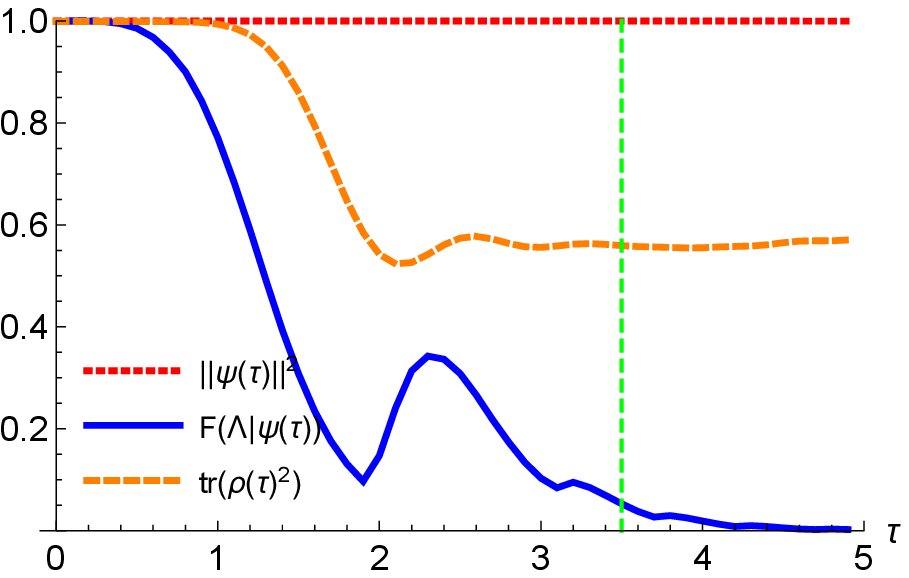}\includegraphics[width=4cm]{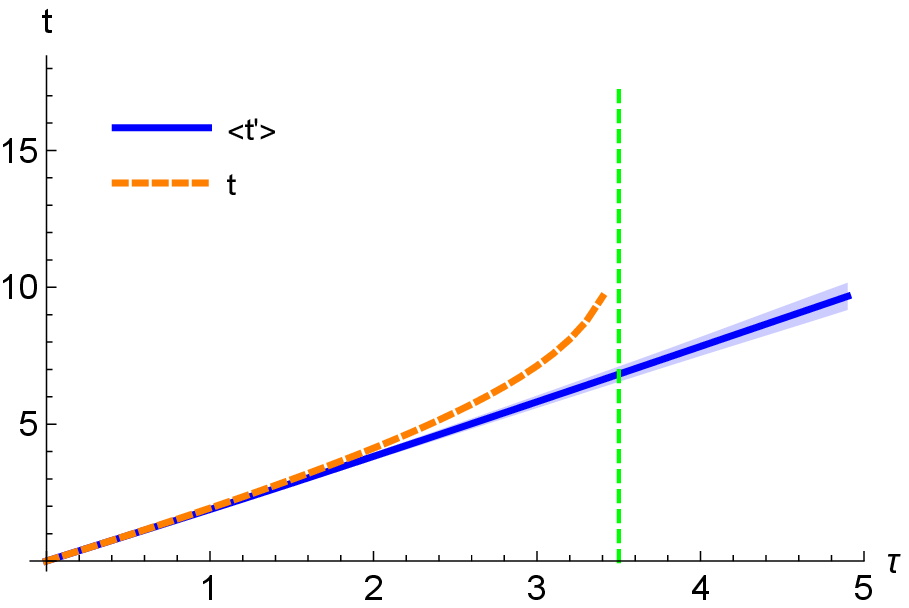}\\
  \includegraphics[width=4cm]{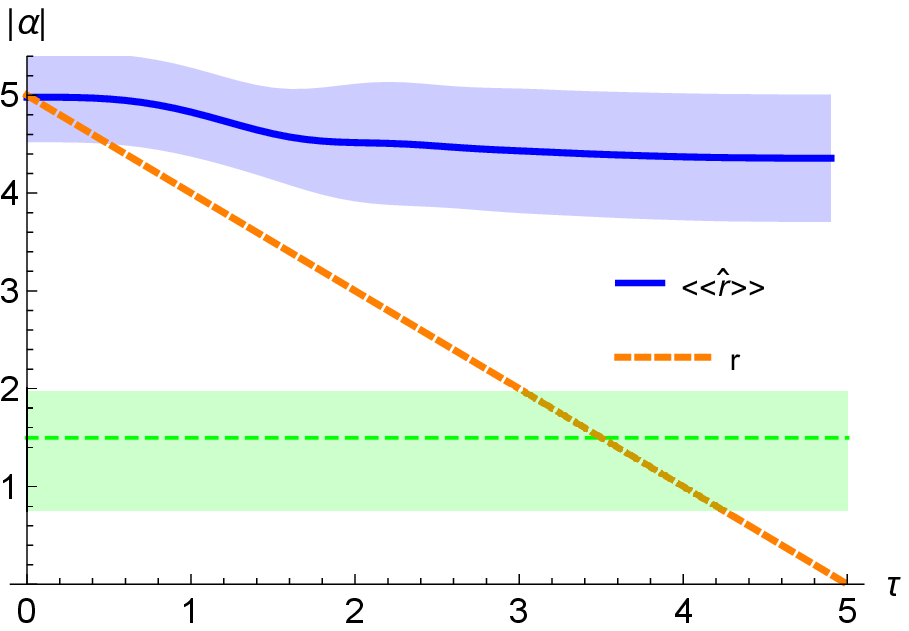}\includegraphics[width=4cm]{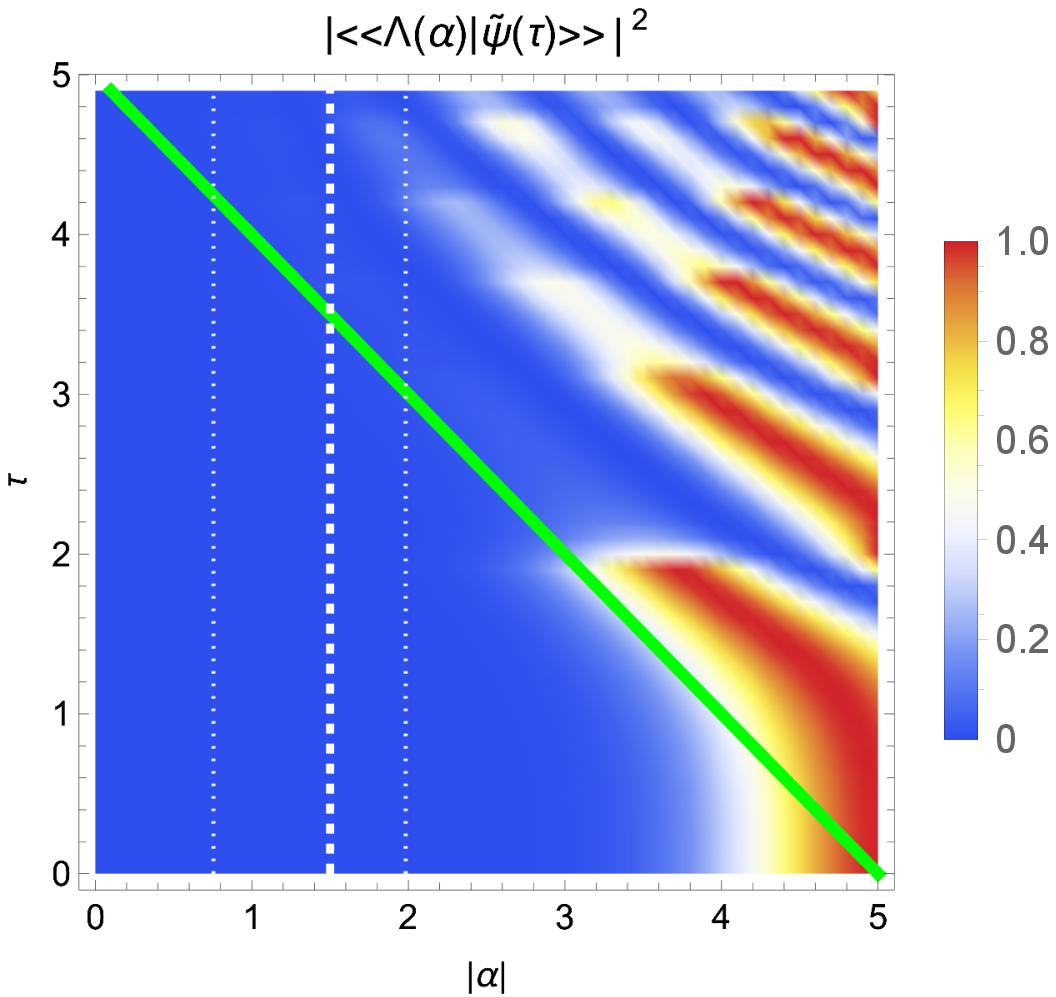}\\
  \caption{\label{uni3} Same as fig.\ref{coheUni} (up left), fig.\ref{rtmoychuteuni} (up right and down left) and fig.\ref{WPuni} (down right), but with $u=1$ ($v(0) \simeq 0.84$).}
\end{figure}

For a large value of $u$, we have a sudden regime for which the end of the string attached on $\mathscr M$ cannot follow the too rapid probe D0-brane, see fig.\ref{uni3}.
Since the particle never arrives at the event horizon, no dissipation occurs. On the contrary, the too rapid drive induces a strong non-adiabaticity in the dynamics and so a rapid fall of the spatial coherence. This one can be associated in the $\alpha$-representation to the oscillations in the driving direction of the spatially coherent part of the state.

\subsection{Geodesic fall}\label{geodesic}
\begin{figure}
  \includegraphics[width=8cm]{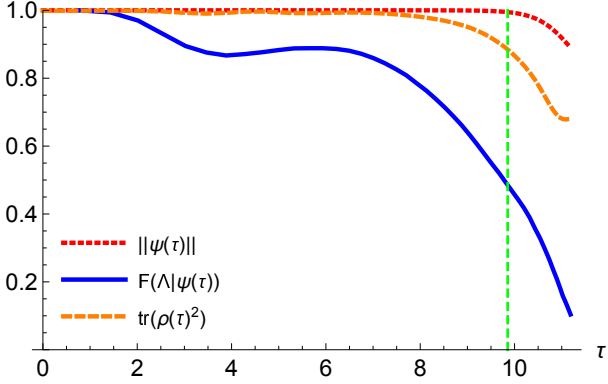}\\
  \caption{\label{coheGeo} Evolutions of the survival probability, of the spatial coherence and of the purity of the fermionic state for a radial geodesic fall of the probe D0-brane with $\epsilon \simeq 0.85$ (corresponding to an initial speed $v(0) \simeq 0.56$). $r_S = 1.5$ (the vertical line corresponds to the passage of the probe D0-brane by the classical event horizon) and $\omega \epsilon'=0$.}
\end{figure}

\begin{figure}
  \includegraphics[width=8cm]{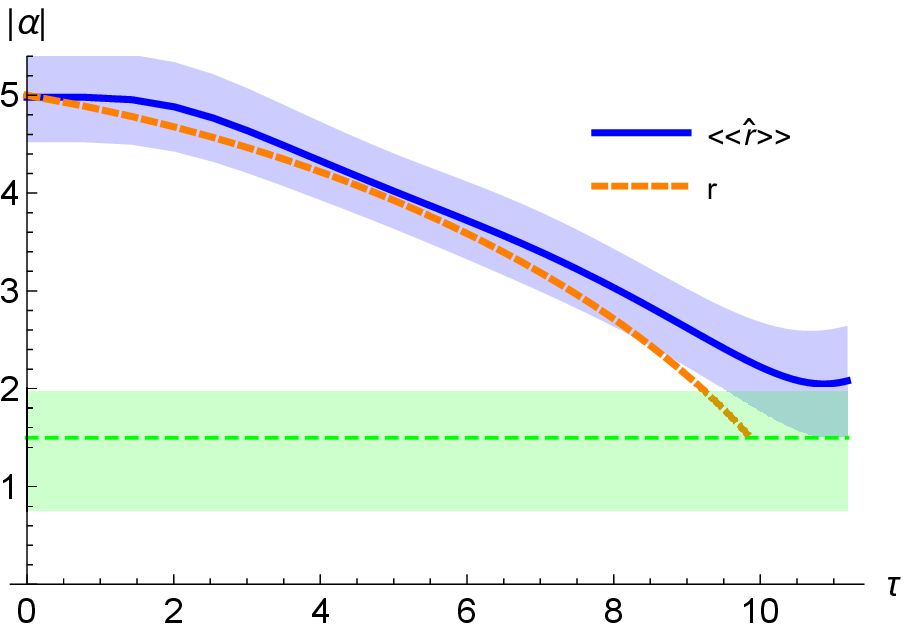}\\
  \includegraphics[width=8cm]{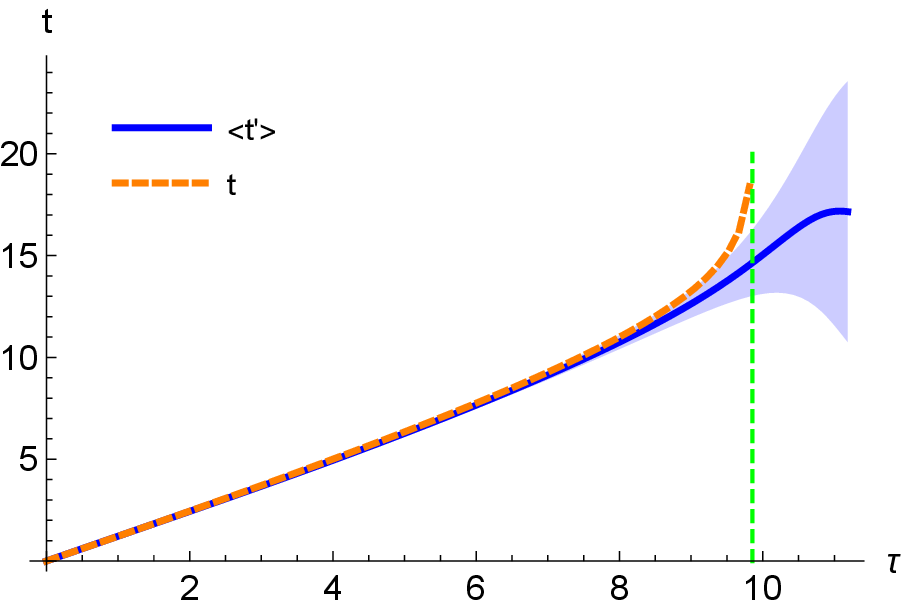}\\
  \caption{\label{rtmoyGeo} $\llangle \hat r \rrangle = \frac{\llangle \psi|\hat r|\psi \rrangle}{\|\psi\|^2}$ (mean value of the radial coordinate of the end of the string attached on $\mathscr M$) and $\langle t'\rangle = \int_0^\tau \frac{\llangle \psi|\hat{\dot t}|\psi \rrangle}{\|\psi\|^2} d\tau$ (mean value of the time of the end of the string measured by the clock of an observer at infinity comoving with the black hole) compared to the radial and time coordinates of the geodetically falling probe D0-brane (with $\epsilon \simeq 0.85$). The light blue clouds correspond to the quantum uncertainties. The parameters are $r_S=1.5$ and $\omega \epsilon'=0$. Up, the dashed horizontal line is the classical event horizon (the light green cloud corresponds to the fuzzy event horizon $\langle r_S \rangle \pm \Delta r_S$); down, the dashed vertical line indicates the time where the probe D0-brane crosses the classical event horizon.}
\end{figure}

\begin{figure}
  \includegraphics[width=8cm]{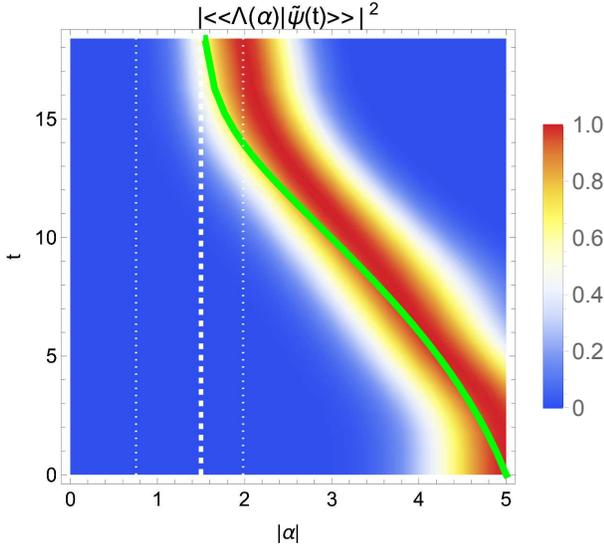}\\
  \caption{\label{WPGeo} Wave packet $\psi(\alpha,t)$ representing the spatially coherent part of the end of the string state compared to the geodesic of the probe D0-brane (green path) with $\epsilon \simeq 0.85$. The white dashed lines represent the classical event horizon $r_S=1.5$ and the fuzzy event horizon $\langle r_S\rangle \pm \Delta r_S$. $\omega \epsilon'=0$.}
\end{figure}

It can be more natural to consider a geodesic fall of the probe D0-brane (considering that the probe brane is not transported but moves under the effect of the gravity). We suppose that the probe D0-brane follows a radial timelike geodesic of the Schwarzschild geometry:
\begin{equation}
  \left\{ \begin{array}{l} \dot r^2 + 1 - \frac{r_S}{r} = \epsilon^2 \\ \left(1-\frac{r_S}{r}\right)\dot t = \epsilon \end{array} \right.
\end{equation}
where $\epsilon$ is the constant of motion (total energy by mass unit). The coherences are drawn fig.\ref{coheGeo}.
As for an uniform transport, we observe decoherence when the probe brane arrives at the event horizon. We have also a small fall of the spatial coherence at the beginning of the dynamics due to a non-adiabatic start. The mean values of $\hat r$ and $t'$ are plotted fig.\ref{rtmoyGeo} and the wave packet $\psi(\alpha,t)$ are plotted fig.\ref{WPGeo}.
The other end of the string follows a path similar to a geodesic close to the probe D0-brane geodesic, except that it seems view the event horizon at $\langle r_S \rangle + \Delta r_S$. The fuzziness of the event horizon seems then be have more consequences for geodesic dynamics.

\subsection{Particle extraction from the black hole}
\begin{figure}
  \includegraphics[width=8cm]{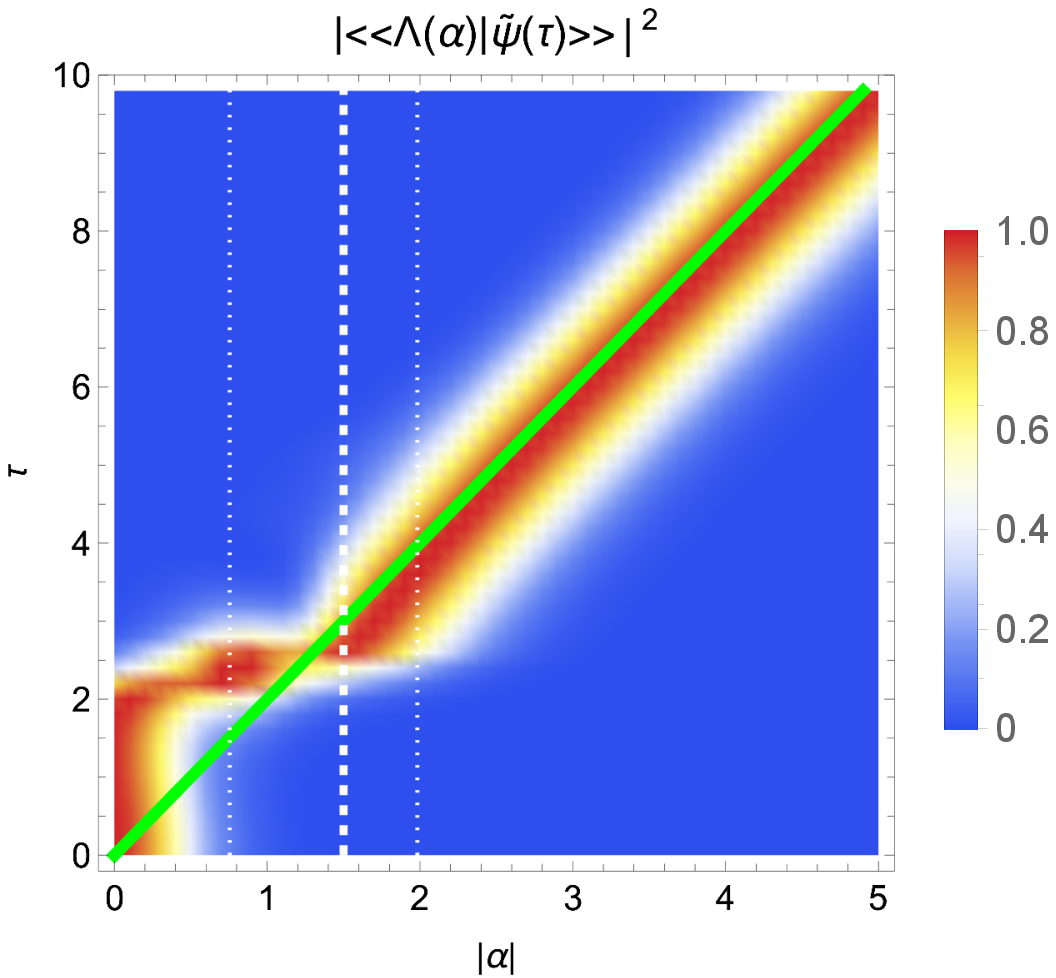}\\
  \includegraphics[width=8cm]{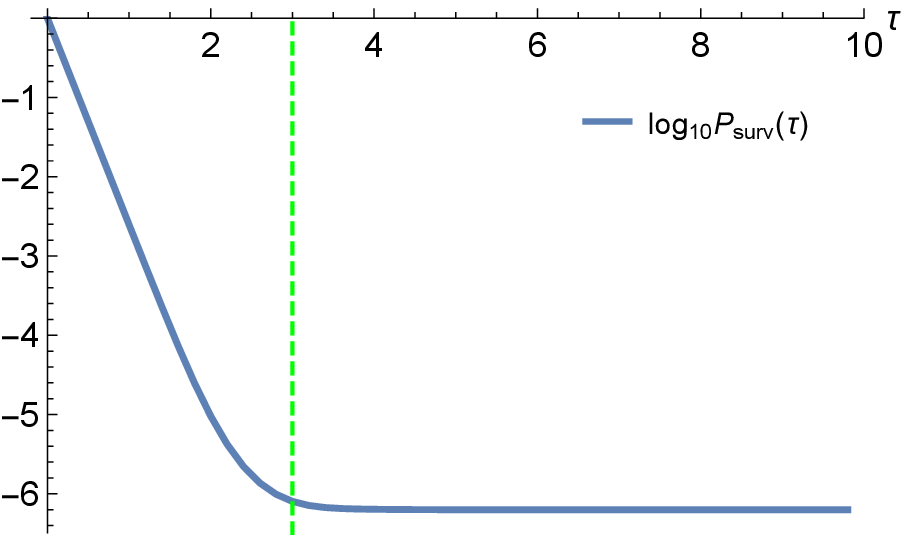}\\
  \caption{\label{extraction} Up: wave packet $\psi(\alpha,\tau)$ representing the spatially coherent part of the end of the string compared to the radial coordinate of the probe D0-brane (green path) with $\dot r=u=0.5$ when we start at the singularity. Down: the associated survival probability $\|\psi(\tau)\|^2$ (in log scale). $\omega \epsilon'=0$.}
\end{figure}

We can try to extract a particle from the black hole by transporting the D0-brane from the singularity to outside of the event horizon with $\dot r=u$ (constant). Fig.\ref{extraction} shows the evolution of the corresponding wave packet $\psi(\alpha,\tau)$ and of the survival probability of the particle.
The particle seems to can be extracted from the quasi-coherent point of view, but as expected the dissipation process by reabsorption by the singularity drastically restricts this possibility, the probability of success being of $\lim_{\tau \to +\infty} p_{surv} \simeq 6.3\times 10^{-7}$ with the parameters used here (in particular $\omega \epsilon'=0$). With massive particle ($\omega \epsilon'\not=0$) the situation is even less favorable because of the well and the barrier at the event horizon induced by $\omega \hat{\dot t}$ (fig.\ref{dotT}). With $\omega \epsilon'=0.5$, we have now $\lim_{\tau \to +\infty} p_{surv} \simeq 1.6\times 10^{-10}$.

\section{White hole and wormhole}
At the construction of the model, we have decide to define $f$ on the Riemann sheet such that $\sqrt{-r} = -\imath \sqrt{r}$ (with $r>0$), in order to the anti-selfadjoint part of $Z$ ($-\imath \Im\, Z$) be a dissipator. If we choose now the other Riemann sheet ($\sqrt{-r} = +\imath \sqrt{r}$), the anti-selfadjoint part of $Z$ ($+\imath \Im\, Z$) becomes a quantum flux generator (it increases the state norm with the time). In place of absorbing particles, the singularity now ``emits particles''. We can then assimilate this case to a toy model of white hole. The Dirac operator for this white hole is simply $\slashed D_\alpha^\dagger$ (with $\slashed D_\alpha$ still defined by eq.(\ref{DiracOp})). This is well the operator governing a quantum dynamics of a time reversed black hole, in accordance with the classical definition of a white hole.\\
Consider the decomposition between selfadjoint and anti-selfadjoint parts of the Dirac operator:
\begin{eqnarray}
  \slashed D_\alpha & = & \underbrace{\left(\begin{array}{cc} \Re\, Z-z(\alpha) & a^+-\bar \alpha \\ a-\alpha & -\Re\, Z+z(\alpha) \end{array} \right)}_{\Re\, \slashed D_\alpha} \nonumber \\
  & & \qquad - \imath \underbrace{\left(\begin{array}{cc} \Im\, Z & 0 \\ 0 & \Im\, Z \end{array} \right)}_{\Im \slashed D}
\end{eqnarray}
We see that $\slashed D_\alpha^\dagger = \Re\, \slashed D_\alpha + \imath \Im\, \slashed D$ whereas $\overline{\slashed D_\alpha} = \Re\, \slashed D_{\bar \alpha} + \imath \Im\, \slashed D$ and so
\begin{equation}
  \slashed D_\alpha^\dagger = \overline{\slashed D_{\bar \alpha}}
\end{equation}
and then
\begin{eqnarray}
  & & \slashed D_\alpha |\Lambda_{BH}(\alpha)\rrangle = \lambda_0(\alpha)|\Lambda_{BH}(\alpha) \rrangle \\
  & \Rightarrow & \overline{\slashed D_{\bar \alpha}} \overline{|\Lambda_{BH}(\bar \alpha)\rrangle} = -\lambda_0(\bar \alpha) \overline{|\Lambda_{BH}(\bar \alpha) \rrangle} \\
  & \Rightarrow & \slashed D_\alpha^\dagger \overline{|\Lambda_{BH}(\bar \alpha)\rrangle} = -\lambda_0(\bar \alpha) \overline{|\Lambda_{BH}(\bar \alpha) \rrangle} \\
  & \Rightarrow &  \slashed D_\alpha^\dagger |\Lambda_{WH}(\alpha)\rrangle = -\lambda_0(\bar \alpha)|\Lambda_{WH}(\alpha) \rrangle 
\end{eqnarray}
with the quasi-coherent state of the white hole defined by
\begin{equation}
  |\Lambda_{WH}(\alpha) \rrangle = \overline{|\Lambda_{BH}(\bar \alpha)\rrangle}
\end{equation}
Note that the duality $|\Lambda_{BH}\rrangle/|\Lambda_{WH}\rrangle$ is the duality between left and right eigenvectors in the non-hermitian quantum dynamics \cite{Moiseyev}. 

\subsection{Quantum transition from black hole to white hole}
\begin{figure}
    \includegraphics[width=8cm]{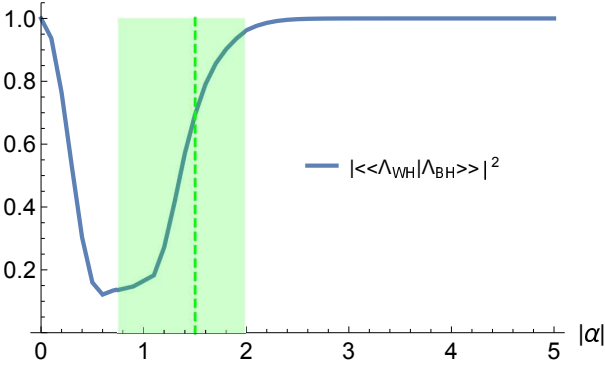}
    \caption{\label{PBHtoWH} Probability of transition from a black hole to a while hole with respect to $|\alpha|$. The vertical dashed line represents the classical event horizon $r_S=1.5$, the light green cloud represents the fuzzy event horizon ($\langle r_S\rangle \pm \Delta r_S$).}
  \end{figure}

\begin{figure}
    \includegraphics[width=8cm]{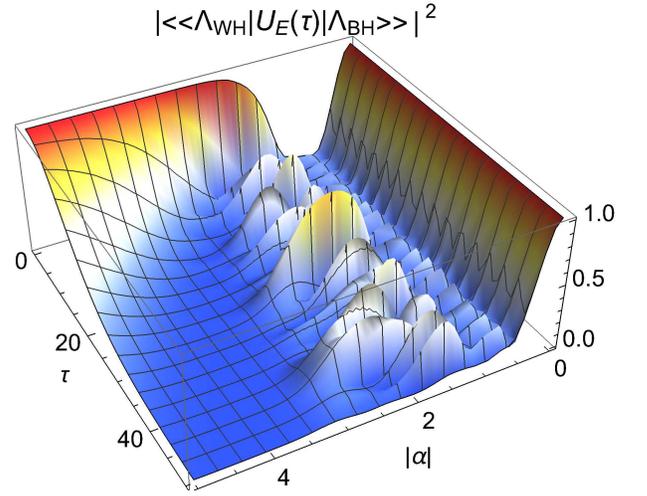}
    \caption{\label{PBHtoWH3D} Time-dependent probability of transition from a black hole to a while hole with respect to $|\alpha|$ and $\tau$ with $r_S=1.5$ and $\omega\epsilon'=1$}
  \end{figure}

It can be possible that a black hole can bounce out its proper event horizon via quantum tunneling into a white hole (see for example \cite{Christodoulou}). It is then interesting to study the probability of transition from a black hole to white hole in our toy model:
  \begin{eqnarray}
    P_{BH\to WH}(\alpha) & = & |\llangle \Lambda_{WH}(\alpha)|\Lambda_{BH}(\alpha)\rrangle|^2 \\
    & = & |\overline{\llangle \Lambda_{BH}(\bar \alpha)}|\Lambda_{BH}(\alpha)\rrangle|^2
  \end{eqnarray}
  This one is plotted fig.\ref{PBHtoWH}.
  It is important to well interpret $P_{BH\to WH}(\alpha)$: this is the probability for an observer comoving with the probe D0-brane at $\alpha$ to see a transition from a black hole fuzzy spacetime to a white hole fuzzy spacetime after a measurement onto the test particle (fermionic string). We note that $P_{BH\to WH}(\alpha) \simeq 1$ for $|\alpha| \gg r_S$. This is consistent with the fact that the Schwarzschild geometry is asymptotically Minkowskian. In the neighorhood of $\alpha$ with large $|\alpha|$, the spacetime is Minkowskian and there is no difference between black and white holes for the observer. This difference can be seen only in neighborhood of the event horizon. Moreover we have no take into account the time effects, by considering that the observer makes measures onto the test fermionic string immediately after its emission/preparation in a quasi-coherent state. Consider the following change of representation:
\begin{equation}
  |\psi \rrangle = U_{E}|\hat \psi \rrangle \text{ with } U_{E}(\tau) = e^{-\imath \omega \hat{\dot t} \tau}
\end{equation}
into eq.(\ref{DiracEeq}):
\begin{equation}
  \imath \frac{\partial|\hat \psi \rrangle}{\partial \tau} = \underbrace{U^{-1}_E(\tau) \slashed D_{\alpha} U_E(\tau)}_{\hat{\slashed D}_{\alpha,\tau}} |\hat \psi(\tau) \rrangle
\end{equation}
We have then a time-dependent eigenvectors of $\hat{\slashed D}_{\alpha,\tau}$ which is $|\hat \Lambda_{BH}(\alpha,\tau)\rrangle = U_E(\tau)|\Lambda_{BH}(\alpha)\rrangle$ and then a time-dependent probability of transition
\begin{equation}
  P_{BH \to WH}(\alpha,\tau) = |\overline{\llangle \Lambda_{BH}(\bar \alpha)}|U_E(\tau)|\Lambda_{BH}(\alpha)\rrangle|^2
\end{equation}
This one is the probability for an observer comoving with the probe D0-brane at $\alpha$ to see a transition from a black hole fuzzy spacetime to a white hole fuzzy spacetime during a measurement after a proper time $\tau$ from the ``preparation'' of the fermionic string in the quasi-coherent state at the position $\alpha$. The time-dependent transition probability is drawn fig.\ref{PBHtoWH3D}.
The probability falls to $0$ for $|\alpha|\gg r_S$ with sufficiently large $\tau$. For short time, the observer cannot have the time to receive information from the event horizon and cannot then distinguish a white hole from a black hole by a local measure onto the test particle. But with $\tau$ sufficiently large, this confusion disappears and it becomes impossible to realize a transition from a black to a white hole by a measurement far from the event horizon. In contrast, the probability of this transition rests possible at the neighborhood of the event horizon (but it is submitted to fast oscillations). 

\subsection{Einstein-Rosen like bridge}
Isolated fuzzy white hole causes a problem in the presented model, because the normalization of the state $|\psi\rrangle$ exponentially increases during the dynamics. As in other approaches of black hole physics, it is more consistent to consider white holes as partners of black holes within wormholes as Einstein-Rosen bridges. The present formalism is not adapted to quantize a Schwarzschild spacetime in Kruskal-Szekeres coordinates for which the extension of a black hole with white hole is natural. But we can propose a modification of our toy model in Schwarzschild coordinates to built an Einstein-Rosen like bridge.\\

We consider two spacetime regions denoted by $\mathcal M_+$ (corresponding to a black hole) and $\mathcal M_-$ (corresponding to a white hole), in which the fuzzy Schwarzschild manifold $\mathscr M$ is embedded, in the sense that $\mathcal M_+ \sqcup_{(0,0)} \mathcal M_-$ (the surgery of the two manifolds gluing at $\alpha=0$) is the new configuration space of the probe D0-brane. Let $(|+\rangle,|-\rangle)$ be the quantum states of occupation of these regions. We consider then now, the Hilbert space $\mathscr H = \mathbb C^2 \otimes \mathbb C^2 \otimes \mathscr F$ (spacetime region states $\otimes$ spin states $\otimes$ noncommutative space wavefunctions). The dissipator $-\imath \Im\, Z$ of the black hole Dirac operator must now make the particle disappear in the region $\mathcal M_+$ to make it reappear in the region $\mathcal M_-$ (and conversely for the generator $+\imath \Im\, Z$ of the white hole Dirac operator). $\pm \imath \Im\, Z$ must be then now couplings between the states $|\pm \rangle$. We propose then the following Dirac operator for the wormhole:
\begin{equation}
  \slashed D_{\alpha}^{\asymp} = \begin{cases} \left(\begin{array}{cc} \Re\, \slashed D_\alpha & \imath \Im\, Z \\ -\imath \Im\, Z & \Re\, \slashed D_0 \end{array} \right) & \text{if } \bullet \in \mathcal M_+ \\
    \left(\begin{array}{cc} \Re\, \slashed D_0 & \imath \Im\, Z \\ -\imath \Im\, Z & \Re\, \slashed D_\alpha \end{array} \right) & \text{if } \bullet \in \mathcal M_- 
  \end{cases}
\end{equation}
$\bullet$ denoting the probe D0-brane. $\slashed D_{\alpha}^{\asymp}$ is self-adjoint because now the quantum flux disappearing in $\mathcal M_+$ reappears in $\mathcal M_-$. The quasi-coherent state is extended as
\begin{equation}
  |\Lambda^\asymp(\alpha) \rrangle = \begin{cases} |+\rangle \otimes|\Lambda_{BH}(\alpha) \rrangle & \text{if } \bullet \in \mathcal M_+ \\ |-\rangle \otimes\overline{|\Lambda_{BH}(\bar \alpha) \rrangle} & \text{if } \bullet \in \mathcal M_- \end{cases}
  \end{equation}

\begin{figure}
  \includegraphics[width=8cm]{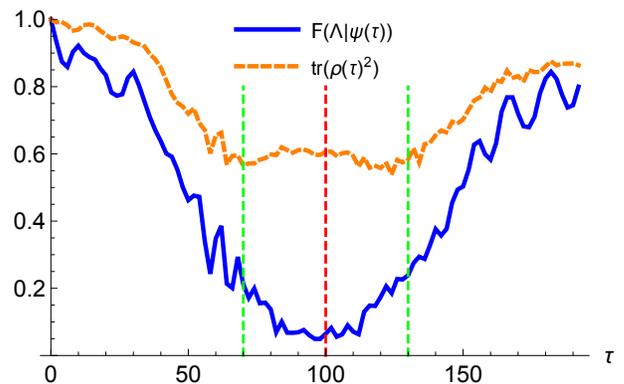}\\
  \caption{\label{coheWorm} Evolutions of the spatial coherence and of the purity of the fermionic state for a radial transport of the probe D0-brane through a wormhole with $|\dot r|=u=0.05$ (corresponding to a initial speed $v(0) \simeq 0.55$). $r_S = 1.5$ (the vertical lines correspond to the passage of the probe D0-brane by the classical event horizons (green) and by the singularity (red)) and $\omega \epsilon'=0$.}
\end{figure}

\begin{figure}
  \includegraphics[width=8cm]{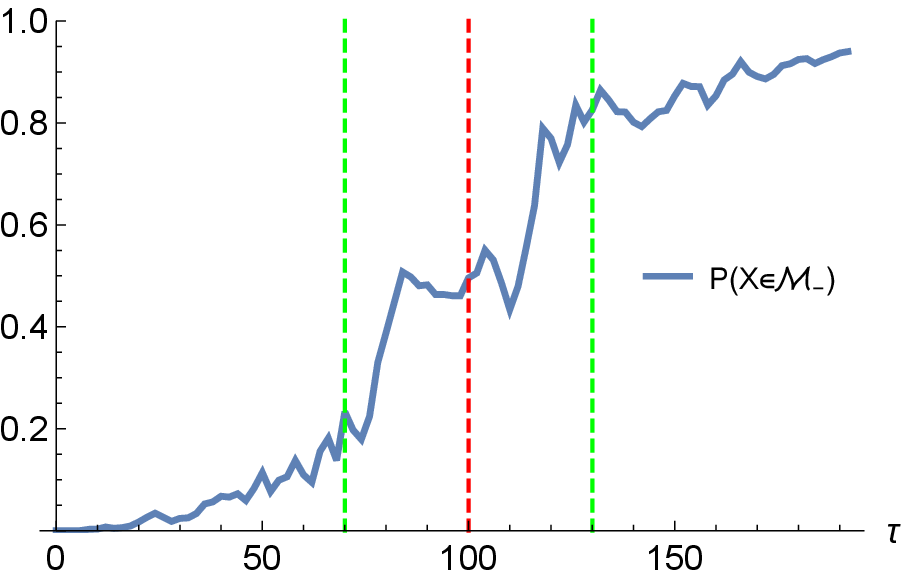}\\
  \includegraphics[width=8cm]{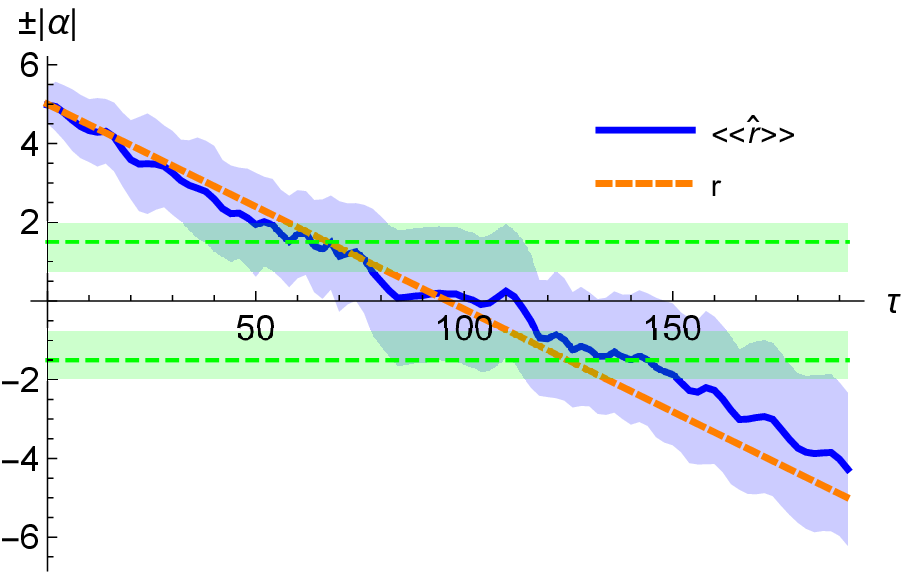}\\
  \includegraphics[width=8cm]{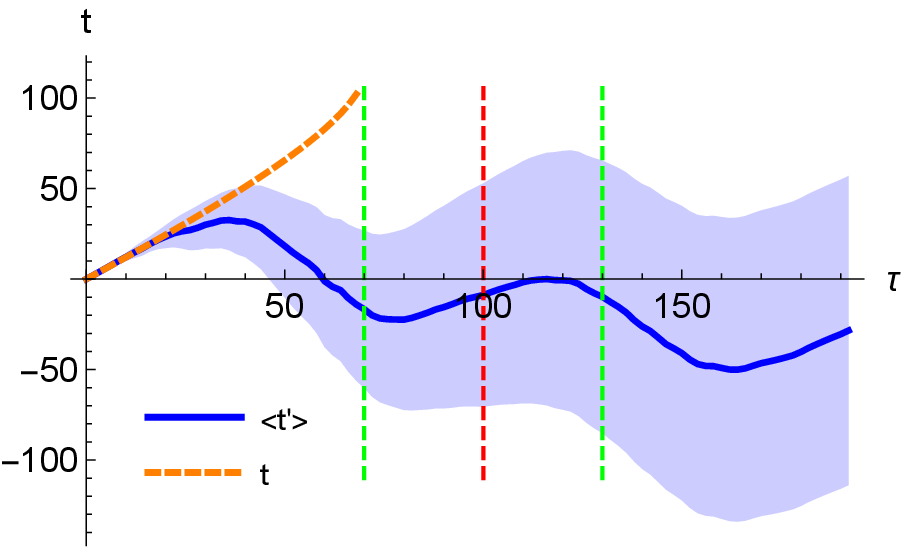}\\
  \caption{\label{rtmoyWorm} Up: probability to find in the spacetime region $\mathcal M_-$ the end of the string attached on $\mathscr M$. Middle: $\llangle \hat r \rrangle = \frac{\llangle \psi|\hat r|\psi \rrangle}{\|\psi\|^2}$ (mean value of the radial coordinate of the end of the string attached on $\mathscr M$) compared to the radial coordinate of the probe D0-brane (the vertical axis is graduated positively for $\mathcal M_+$ and negatively for $\mathcal M_-$). Down: $\langle t'\rangle = \int_0^\tau \frac{\llangle \psi|\hat{\dot t}|\psi \rrangle}{\|\psi\|^2} d\tau$ (mean value of the time of the end of the string measured by the clock of an observer at infinity in $\mathcal M_+$ comoving with the black hole) compared to the time coordinate of the probe D0-brane. The light blue clouds correspond to the quantum uncertainties. The parameters are $|\dot r|=u=0.05$,   $r_S=1.5$ and $\omega \epsilon'=0$. The dashed horizontal lines are the classical event horizons (the light green cloud corresponds to fuzzy event horizon $\langle r_S \rangle \pm \Delta r_S$); the dashed vertical line indicates the time where the probe D0-brane crosses a classical event horizon or the singularity.}
\end{figure}

\begin{figure}
  \includegraphics[width=8cm]{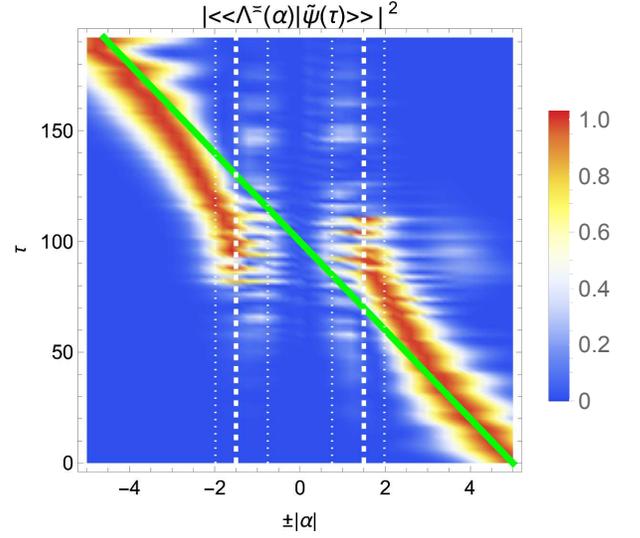}
  \caption{\label{WPWorm} Wave packet $\psi(\alpha,\tau)$ representing the spatially coherent part of the end of the string state compared to the radial coordinate of the probe D0-brane (green path) passing through a wormhole when $|\dot r|=u=0.05$. The white dashed lines represent the classical event horizons $r_S=1.5$ and the fuzzy event horizons $\langle r_S\rangle \pm \Delta r_S$. $\omega \epsilon'=0$. The horizontal axis is graduated positively for $\mathcal M_+$ and negatively for $\mathcal M_-$.}
\end{figure}

\begin{figure}
  \includegraphics[width=4cm]{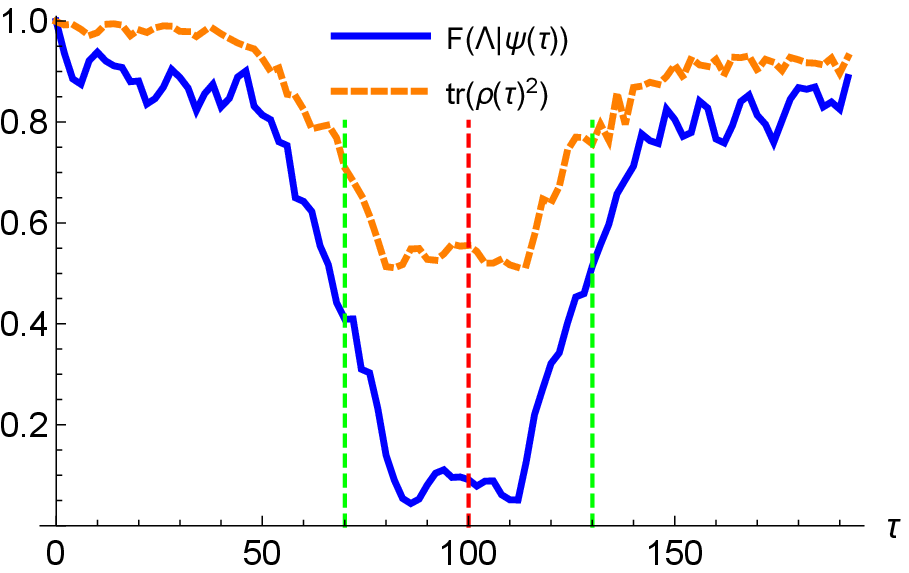}\includegraphics[width=4cm]{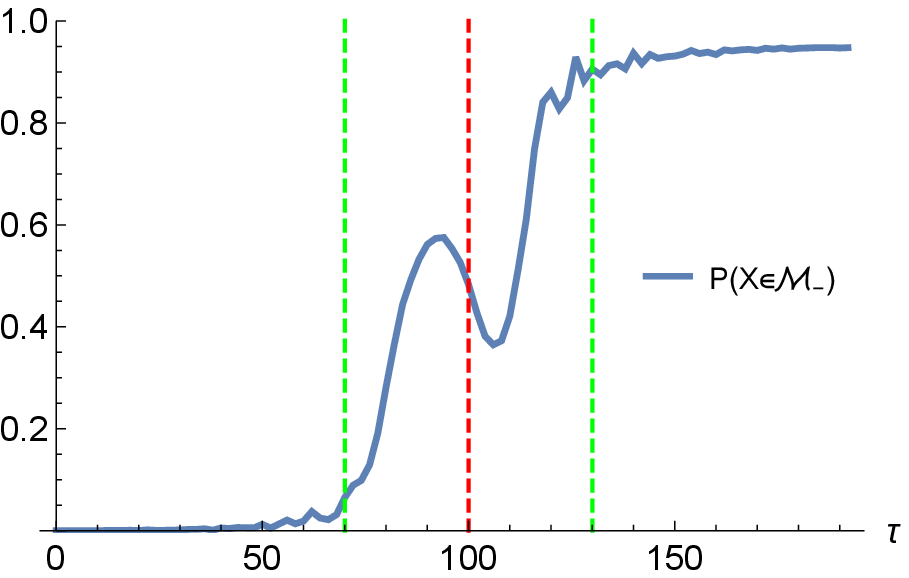}\\
  \includegraphics[width=4cm]{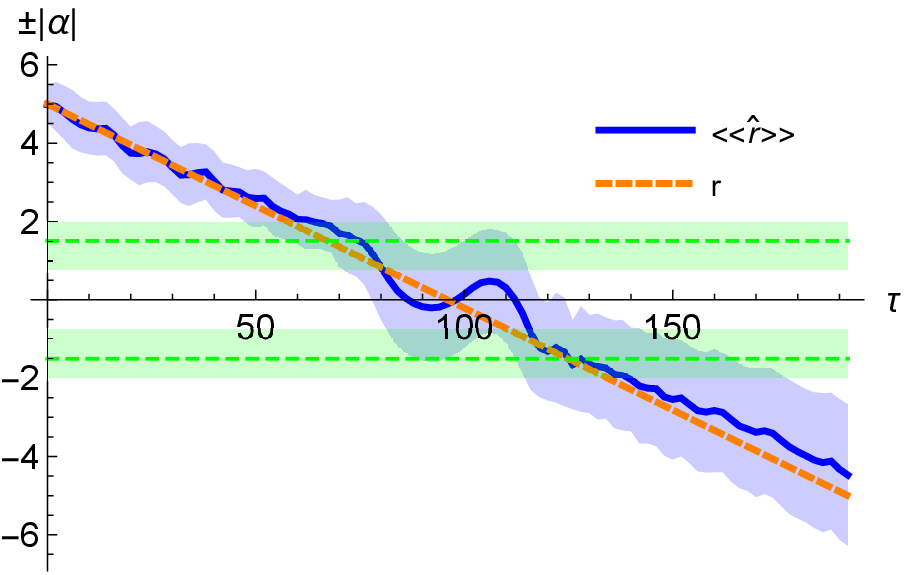}\includegraphics[width=4cm]{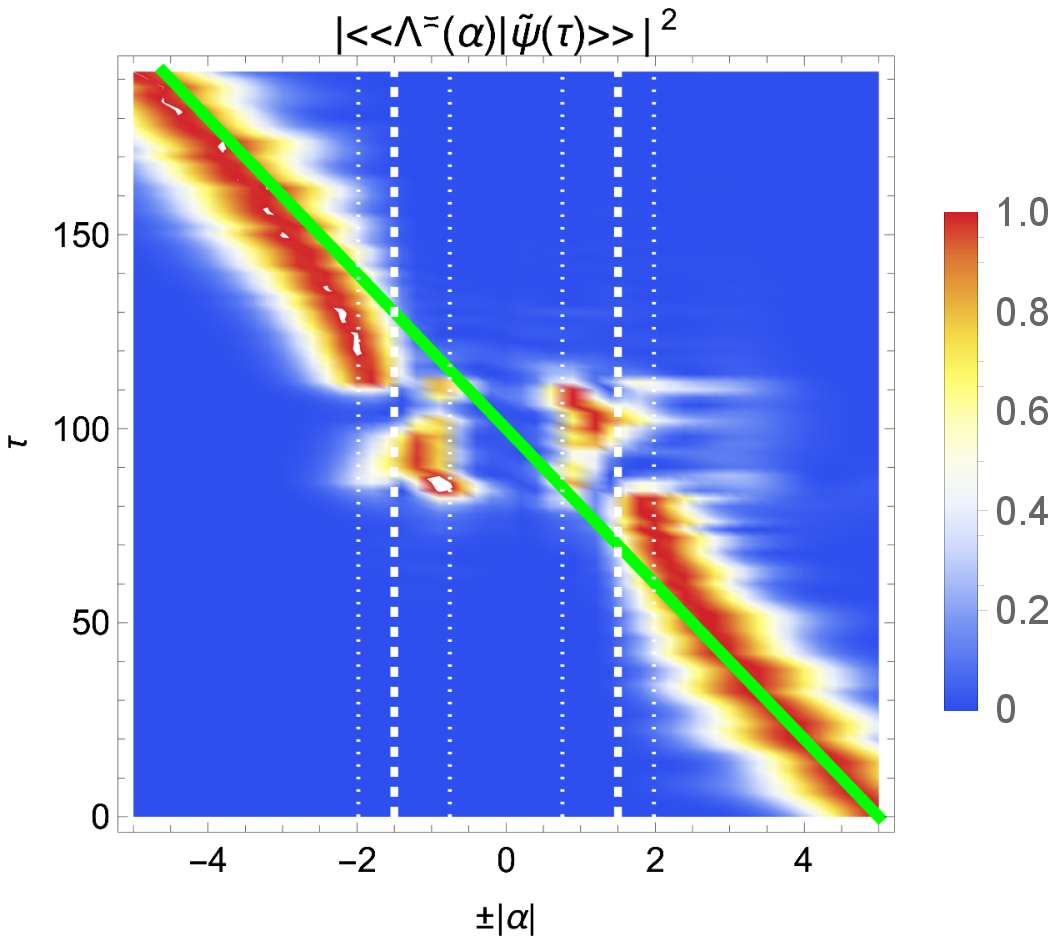}\\
  \caption{\label{Worm2} Same as fig.\ref{coheWorm} (up left), fig.\ref{rtmoyWorm} (up right and down left) and fig.\ref{WPWorm} (down right), but with $\omega \epsilon'=0.09$.}
\end{figure}

We consider an uniform radial transport of the D0-brane traveling through $\mathcal M_+$ to $\mathcal M_-$, with $|\dot r|=u$ (constant). For a small value of $u$, the coherences are plotted fig.\ref{coheWorm}.
The coherences fall during the transport on $\mathcal M_+$ but increase during the transport on $\mathcal M_-$, in accordance with the ``anti-symmetric'' role of the white hole. The mean observables during the transport are drawn fig.\ref{rtmoyWorm}.
The particle passes through the wormhole, the process starting with a small probability which strongly increases when the D0-brane is between the two event horizons. The mean value of $\hat r$ follows the probe D0-brane. $\langle t'\rangle$ is meaningless in that case, since it corresponds to a measure by a distant observer in $\mathcal M_+$ which formally counts negatively the elapsed time between the two horizons. But the average value or $\hat r$ hides an important phenomenon which can be revealed by drawing the wave packet $\psi(\alpha,\tau) = \llangle \Lambda^\asymp(\alpha)|\tilde \psi(\tau)\rrangle$, fig.\ref{WPWorm}.
The quasi-coherent wave packet does not come in the wormhole. As observed for the fall into a quantum black hole, the wave packet bounces on the event horizon and stagnates in its neighborhood. The part which is transmitted by the event horizon (which not absorbed now), directly reappears in the neighborhood of the second horizon. The quasi-coherent wave packet does not travel through the wormhole but passes through it by quantum tunneling. This is in accordance with the fact that the Einstein-Rosen bridges are classically non-traversable. With large values of $\omega \epsilon'$, the tunneling process is tone off and the particle remains stay stuck in front of the first event horizon (for $\llangle \hat r \rrangle$ and for the wave packet) because of the large barrier introduced by $\omega \hat{\dot t}$ (as in the previous examples, if $\omega \epsilon'> M$, the wormhole cannot transfer a particle bigger than itself). In contrast, a small non-zero value of $\omega \epsilon'$ increases the quantum tunneling through the wormhole, see fig.\ref{Worm2}.
This can be understand by regarding fig.\ref{dotT}. With reasonable value of $\omega \epsilon'$ the wave packet can overcome the small barrier ahead the event horizon but can be trapped by the deep well behind the horizon. The wave packet can be then easily passes by tunneling from the deep well of the first horizon to the one of the second. This can be confirmed by fig.\ref{Worm2} down right, where we see wave packet trapped behind the event horizons.

\section{Conclusion}
A toy model representing the quantization of a $(r,\theta)$ Schwarzschild slice is proposed as a fuzzy manifold. This model permits to describe a fuzzy event horizon by a purely quantum model (in contrast with previous works as \cite{Nicolini, Banerjee} where a classical geometry is computed from a delocalized singularity viewed as a smeared mass). The Dirac operator is not self-adjoint, the dissipator being interpreted as the effect of particle absorption by the black hole singularity. The geometry defined by the quasi-coherent states is consistent with the Schwarzschild geometry with decoherence effects occurring in the neighborhood of the fuzzy event horizon. The model can be modified to describe a quantum Einstein-Rosen like bridge in Schwarzschild coordinate. In this model, a particle do not travel through the wormhole interior but crosses the bridge by quantum tunneling.\\
We have considered only a space slice (with Schwarzschild coordinates) embedded in $\mathbb R^3$, with a simple fuzzy manifold obtained by deformation of the noncommutative plane. Ref. \cite{Blaschke} considers a 4D spacetime embedded in $\mathbb R^7$ (with Eddington-Finkelstein coordinates) but with the noncommutative manifold obtained as a ``semiclassical manifold'' by using a Moyal-type star product. An interesting question is the possibility to find intermediate models of noncommutative manifold related to the Schwarzschild geometry. In particular, Since quantum event horizons can be described by fuzzy spheres \cite{Dolan,Iizuka,Argota}, a future work could consist to built a composite model from the fuzzy sphere and the fuzzy Schwarzschild slice. Another question is the signature of the noncommutativity at the emergent geometry in the quasi-coherent representation. In ref.\cite{Steinacker2} it is proved that at the semi-classical limit, a non zero torsion is a signature of the noncommutativity of the quantum spacetime. Moreover a general framework based on the geometric phases of the adiabatic transport to define such non zero torsion with fuzzy geometries has proposed in \cite{Viennot}. This method is not applicable here because of the strong non-adiabaticity of the model in the neighborhood of the fuzzy event horizon. Maybe with a fuzzy manifold obtained by quantization of a different coordinate system regular at the horizon (as Kruskal-Szekeres coordinates) could avoid this problem and permits to study a torsion induced by the spacetime noncommutativity.

\appendix
\section{The noncommutative plane}\label{NCP}
The noncommutative plane is defined by eq.(\ref{NCPlane}). The associated Dirac operator is then
\begin{equation}
  \slashed D_\alpha = \left(\begin{array}{cc} 0 & a^+-\bar \alpha \\ a - \alpha & 0 \end{array} \right)
\end{equation}
The quasi-coherent state, solution of $\slashed D_\alpha |\lambda_0(\alpha) \rrangle = 0$, is simply
\begin{equation}
  |\lambda_0(\alpha) \rrangle = |0 \rangle \otimes |\alpha \rangle
\end{equation}
where
\begin{equation}
  |\alpha \rangle = e^{-|\alpha|^2/2} \sum_{n=0}^{+\infty} \frac{\alpha^n}{\sqrt{n!}} |n\rangle
\end{equation}
is the Perelomov coherent state \cite{Perelomov} of the quantum harmonic oscillator algebra: $a|\alpha\rangle = \alpha|\alpha \rangle$. Coherent states are not orthogonal:
\begin{equation}
  |\langle \beta|\alpha \rangle|^2 = e^{-|\alpha-\beta|^2}
\end{equation}
and correspond then to gaussian localizations around $\alpha \in \mathbb C$ minimizing the quantum uncertainties.\\
More generally, we have (for $n>0$)
\begin{eqnarray}
  & & \slashed D_\alpha |\lambda_{n\pm}(\alpha) \rrangle = \pm \sqrt n |\lambda_{n\pm}(\alpha) \rrangle \\
  & & |\lambda_{n\pm}(\alpha) \rrangle = \frac{1}{\sqrt 2}(|0\rangle \otimes |n\rangle_\alpha \pm |1\rangle \otimes |n-1\rangle_\alpha)
\end{eqnarray}
where
\begin{equation}
  |n\rangle_\alpha = \frac{(a^+-\bar \alpha)^n}{\sqrt{n!}} |\alpha \rangle
\end{equation}
(with the identification $|\alpha\rangle = |0\rangle_\alpha$). $(|n\rangle_\alpha)_{n \in \mathbb N}$ constitutes an orthonormal basis of the Fock space $\mathscr F$ as the canonical basis of the representation of the quantum harmonic oscillator operators $b_\alpha = a-\alpha$ and $b_\alpha^+ = a^+ - \bar \alpha$ ($[b_\alpha,b_\alpha^+]=1$): $b^+_\alpha |n\rangle_\alpha = \sqrt{n+1}|n+1\rangle_\alpha$ and $b_\alpha |n\rangle_\alpha = \sqrt n |n-1\rangle_\alpha$. We can then view $(|n\rangle_\alpha)_{n \in \mathbb N}$ as the local basis of $\mathscr F$ (``local'' in the meaning of the $\alpha$-representation).\\

Let $r(\tau) = r_0-\gamma v\tau$ (with $\gamma = \frac{1}{\sqrt{1-v^2}}$) be a geodesic transport of the probe D0-brane ($\alpha(\tau)=r(\tau) e^{\imath \theta}$). We consider the Dirac equation:
\begin{eqnarray}
  & & \imath |\dot \psi \rrangle = \slashed D_{\alpha(\tau)} |\psi(\tau) \rrangle \\
  & & |\psi(0)\rrangle = |\lambda_0(\alpha(0))\rrangle
\end{eqnarray}
where the dot denotes the derivative with respect to the proper time $\tau$. Since $\dot {\slashed D_\alpha} = \dot r {\scriptstyle \left(\begin{array}{cc} 0 & e^{-\imath \theta} \\ e^{\imath \theta} & 0 \end{array}\right)}$, the dynamical behavior is governed by $|\dot r|=\gamma v$:
\begin{itemize}
\item For $v \ll 1 \iff |\dot r|\ll 1$, the dynamics is adiabatic and then
  \begin{equation}
    |\psi(\tau) \rrangle \simeq e^{\imath \varphi(\tau)}|\lambda_0(\alpha(\tau))\rrangle
  \end{equation}
  where $e^{\imath \varphi(\tau)}$ is the dynamical and geometric phases. The mean value of the ``wave packet'' $\psi(\alpha,\tau) = \llangle 0,\alpha|\psi(\tau)\rrangle$ follows then the classical geodesic of the probe D0-brane and the spatial coherence (fidelity to a quasi-coherent state) remains 1 during the whole dynamics. Numerical simulations show that this regime is valid until $v \simeq 0.3$ ($\gamma v \simeq 0.3$).
\item For $v \sim 1 \iff |\dot r| \sim +\infty$, the dynamics is sudden and then
  \begin{equation}
    |\psi(\tau) \rrangle \approx |\lambda_0(\alpha(0)) \rrangle
  \end{equation}
  The wave packet remains at $\alpha(0)$ with a strong loss of coherence (the quantum state cannot follow the too rapid D0-brane). Numerical simulations show that this regime starts to dominate from $v \simeq 0.6$ ($\gamma v \simeq 0.75$).
\end{itemize}
With $v \in [0.3,0.6]$ the dynamics is a mix of the adiabatic and sudden regimes.

\section{About numerical computations}\label{nummeth}
\begin{figure}
  \includegraphics[width=8cm]{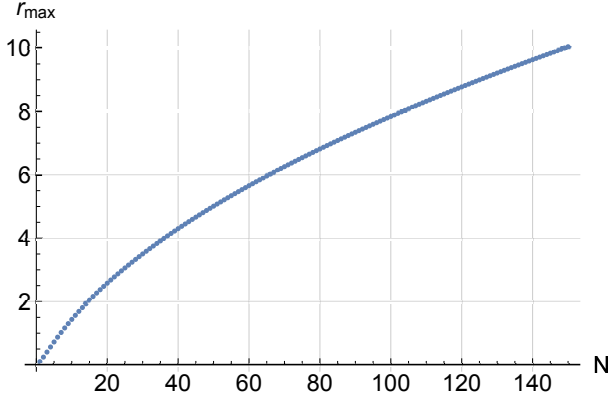}\\
  \caption{\label{numtrunc} Maximal radius in the complex plane for which the coherent state equation is valid (with an accuracy $\epsilon = 10^{-2}$) with respect to the truncation value $N$.}
\end{figure}

In numerical computations we work with the truncated Fock space $\mathscr F_N = \Lin\left(|n\rangle, n\in\{0,...,N\}\right)$. The operators of the quantum harmonic oscillator algebra are then represented by
\begin{eqnarray}
  a_N & = & \sum_{n=1}^N \sqrt n |n-1 \rangle \langle n| \\
  a_N^+ & = & \sum_{n=0}^{N-1} \sqrt{n+1}|n+1\rangle \langle n|
\end{eqnarray}
and the truncated coherent state is
\begin{equation}
  |\alpha \rangle_N = \mathcal N_\alpha \sum_{n=0}^N \frac{\alpha^n}{\sqrt{n!}} |n\rangle
\end{equation}
where $\mathcal N_\alpha = \left(\sum_{n=0}^N |\alpha|^{2n}/n!\right)^{-1/2}$ is the normalization factor. It is an approximate eigenvector of $a_N$:
\begin{equation}
 \forall |\alpha|<r_{max}, \quad  \| (a_N-\alpha)|\alpha\rangle_N \| \leq \epsilon
\end{equation}
The value of $r_{max}$ (maximal value for which the numerical representation is correct) depends on the truncation value $N$ and the expected accuracy $\epsilon$. Fig. \ref{numtrunc} shows this relation.
In the computations presented in this paper, we have chosen $N=50$ involving than $r_{max} = 5$ (with $\epsilon = 10^{-2}$).\\

To solve the equation $\slashed D_\alpha |\Lambda(\alpha)\rrangle = \lambda_0(\alpha)|\Lambda(\alpha)\rrangle$ with $\Re \lambda_0=0$, we consider a lattice $\mathfrak X = \{r_i e^{\imath \theta_j}, r_i = i \frac{r_{max}}{N_r} , \theta_j = \frac{2\pi j}{N_\theta} \}_{i=0,...,N_r; j=0,...,N_{\theta}-1}$ of the disk $|\alpha|\leq r_{max}$. We consider also a partition $\mathfrak X_z = \{k \frac{f(r_{max})}{N_z}\}_{k=0,...,N_z}$. For each point $(\alpha_{i,j},z_k) \in \mathfrak X \times \mathfrak X_z$, we diagonalize $\slashed D_{\alpha_{i,j},z_k,N}$ (Dirac operator built with the truncated creation and annihilation operators) and we select the eigenvalue $\lambda_{ijk}$ with minimal real part. $z_{k_{ij}}$ such that $\Re\, \lambda_{ijk_{ij}} = \min_k \Re\, \lambda_{ijk}$, is a first approximation of $z(\alpha_{ij})$. Next we improve the approximation by dichotomy: we compute $\lambda_{i,j,k_{ij}\pm\frac{1}{2}}$ (with $z_{k_{ij}\pm\frac{1}{2}} = z_{k_{ij}} \pm \frac{f(r_{max})}{2N_z}$) and redefine $z_{k_{ij}}$ with $k_{ij}$ corresponding to the minimal value of $\{\Re\, \lambda_{i,j,k_{ij}}, \Re\, \lambda_{i,j,k_{ij}\pm\frac{1}{2}}\}$. And we iterate the process until we find an eigenvalue with real part equal to zero with the expected accuracy ($\epsilon = 10^{-2}$ in the computations presented in this paper). For computations needed only results on a single radial axis ($\theta=0$ for example), we can only consider a 1D lattice $\mathfrak X_r = \{r_i=i\frac{r_{max}}{N_r}\}_{i=0,...,N_r}$.

\end{document}